\documentclass[showpacs,twocolumn,preprintnumbers,showkeys,superscriptaddress,amsmath,amssymb,nofootinbib]{revtex4}
\usepackage{pstricks,pst-coil}
\usepackage{color}
\usepackage{amsmath}
\usepackage{amssymb}
\usepackage[utf8]{inputenc}
\usepackage{amsfonts}
\usepackage{bm}
\usepackage{bbold}
\usepackage{graphics}
\usepackage{graphicx}

\newcommand{\be}{\begin{equation}}
\newcommand{\ee}{\end{equation}}
\newcommand{\ba}{\begin{eqnarray}}
\newcommand{\ea}{\end{eqnarray}}

\def\ni{\noindent}

\begin{document}

%\begin{center}

%\title{A gravitation model in the Doplicher-Fredenhagen-Roberts noncommutative space-time }
\title{\Large Dispersion Relations in Non-Linear Electrodynamics and the Kinematics of the Compton Effect in a Magnetic Background}

%\bigskip
%\bigskip

\author{M. J. Neves} \email{mariojr@ufrrj.br}
%\affiliation{Department of Physics and Astronomy, University of Alabama, Tuscaloosa, Alabama 35487, USA}
\affiliation{Departamento de F\'{i}sica, Universidade Federal Rural do Rio de Janeiro, BR 465-07, 23890-971, Serop\'edica, RJ, Brazil}

\author{Jorge B. de Oliveira}\email{jorgebernardo998@gmail.com}
\affiliation{Departamento de F\'{i}sica, Universidade Federal Rural do Rio de Janeiro, BR 465-07, 23890-971, Serop\'edica, RJ, Brazil}

\author{ L. P. R. Ospedal } \email{leoopr@cbpf.br}
\affiliation{Centro Brasileiro de Pesquisas F\'isicas, Rua Dr. Xavier Sigaud
150, Urca, Rio de Janeiro, Brazil, CEP 22290-180}

\author{J. A. Helay\"el-Neto}\email{helayel@cbpf.br}
\affiliation{Centro Brasileiro de Pesquisas F\'isicas, Rua Dr. Xavier Sigaud
150, Urca, Rio de Janeiro, Brazil, CEP 22290-180}

%%%%%%%%%%%%%%%%%%%%%%%%%%%%%%%%%%%%%%%%%%%%%%%%%%%%%%%%%%%%%%%%%%%%%%%%%%%%%%%%%%%%%%%%%%%%%%%%%%%%%%%%%%%%%%%%%%%%%%%%%%%

\date{\today}

\begin{abstract}
\ni

Non-linear electrodynamic models are re-assessed in this paper to pursue an investigation of the kinematics of the Compton effect
in a magnetic background. Before considering specific models, we start off by presenting a general non-linear Lagrangian built up
in terms of the most general Lorentz- and gauge-invariant combinations of the electric and magnetic fields. The extended Maxwell-like equations
and the energy-momentum tensor conservation are presented and discussed in their generality. We next expand the fields around a uniform and time-independent electric and magnetic backgrounds up to second order in the propagating wave, and compute dispersion relations which account for the effect of the external fields. We obtain thereby the refraction index and the group velocity for the propagating radiation in different situations. In particular, we focus on the kinematics of the Compton effect in presence of external magnetic fields. This yields constraints that relate the derivatives of the general Lagrangian with respect to the field invariants and the magnetic background under consideration. We carry out our inspection by focusing on some specific non-linear electrodynamic effective models: Hoffmann-Infeld, Euler-Heisenberg, generalized Born-Infeld and Logarithmic.

\end{abstract}



\maketitle

\pagestyle{myheadings}
\markright{DRs in Non-Linear Electrodynamics and the Kinematics of the Compton Effect in a Magnetic Background}

%%%%%%%%%%%%%%%%%%%%%%%%%%%%%%%%%%%%%%%%%%%%%%%%%%%%%%%%%%%%%%%%%%%%%%%%%%%%%%%%%%%%%%%%%%%%%%%%%%%%%%%%%%%%%%%%%%%%%%%%%%%%%%%%%%%%%%%%%%%%%%%%%%%%%%%%%%%%%%%%%%%%%%%%%%%%%%%%%%%%%%%%%%%%%%
%\newpage

\section{Introduction}
Maxwell Electrodynamics is a highly successful theory to describe properties of the electromagnetic interaction at
both the classical and the quantum length scales. Photon-photon scattering in Quantum Electrodynamics (QED)
motivates the study of non-linear extensions of the Maxwell Electrodynamics (MED), and phenomena like vacuum birefringence
and vacuum dichroism may be a guide to also inspect the consistency of non-linear extensions of MED  \cite{Adler71,Constantini71,
Biswas07,Seco09,Seco07,Seco08,Kruglov07}. We cite here the works by Plebanski, Boillat, Bialynicka-Birula and Bialynicki-Birula as excellent articles for all those readers who wish to be introduced to the issue of non-linear extensions of MED \cite{Plebanski,Boillat,Birula1,Birula2}. Different scenarios of the latter introduce new effects into the photon-photon scattering process. To be more specific, we quote scenarios such as Born-Infeld Electrodynamics, models with milli-charged particles and models of axion-like particles that interact topologically with hidden-photons \cite{Born,Davila,Ellis,Gies,Masso,GaeteMPLA,GaeteJPA,Arias}. The non-linear Born-Infeld Electrodynamics was originally introduced to remove the singularity of the electric field of point-like charges on their space position. Nowadays, Born-Infeld effective actions emerge in diverse scenarios, like superstring theory, quantum-gravitational models and theories with magnetic monopoles \cite{Fradkin,Pope,Banerjee,Mann,Garcia,NiauEPJC,NiauPRD}. Furthermore, a number of new phenomena in Cosmology and black-hole physics have been reported in connection with non-linear extensions of electrodynamic systems coupled to gravity \cite{Hendi,Pan,Olea,Halisoy,Lopez,Aiello,Balart}.
The introduction of external backgrounds in field-theoretic models is an old procedure to reproduce effects of the vacuum polarization phenomenon. If the external electromagnetic fields are strong enough (as compared to the so-called Schwinger critical electric and magnetic fields), they can induce the creation of real particle-antiparticle pairs in the non-trivial quantum vacuum, such as, the electron-positron pairs of QED vacuum \cite{Dirac,Schwinger}. Non-linear extensions of electrodynamics in connection with external electromagnetic fields are able to describe effects of the vacuum structure on waves that propagate in empty space, as already pointed out in the previous paragraph, through the observation of birefringence and vacuum dichroism phenomena \cite{Liao,Valle1,Valle2,Potting,Baier}. A well-known case of non-linear extension that stems from the vacuum polarization is the Euler-Heisenberg Lagrangian \cite{EH}, which is an effective photonic model with higher powers in the electric and magnetic fields, attained upon integration over the quantum effects of the (virtual) electron-positron pairs.  We would like to point out here the interesting paper of Reference \cite{GiesJHEP}, where the EH model is studied beyond the 1-loop level in a great deal of details. Let us also recall that, in 1961, Franken et al. opened up the field of non-linear optics with the celebrated experiment in which they successfully measured the second-harmonic (optical) generation \cite{Franken}.
Motivations that strongly justify the renewed interest in non-linear extensions of MED are also coming from the recent high-intensity LASERs, which are the best devices to test both Classical and Quantum Electrodynamics in the strong-field regime, whose (field) scales are fixed by the critical intensity $I_{crit} \sim 4.6 \times 10^{29} \, \mbox{W} \cdot \mbox{cm}^{-2}$. The Station of Extreme Light (SEL), the Europe's Extreme Light Infrastructure (ELI Project) and the ExaWatt Center for Extreme Light Studies (XCELS) shall, in a close future, provide the facilities to pursue a very detailed inspection of electromagnetic non-linearities and allow a real dive into the structure of the quantum vacuum. Powers of these extremely-intense LASERs are expected to reach $10^{2}$ PW, hopefully getting even to the Exa-W scale in a couple of years.  We bring to the reader's attention that high-intensity LASER experiments designed to inspect non-linearity effects in QED are presented and discussed in \cite{Felix,Lundin,Battesti,King}.
Non-linearity means corrections to the Maxwell electrodynamics that, in general, depend on the two Lorentz- and gauge-invariant quantities,
${\cal F}=-F_{\mu\nu}^2/4$ and ${\cal G}=-F_{\mu\nu}\widetilde{F}^{\mu\nu}/4$. In many examples, the Lagrangian density of the non-linear model
depends exclusively on powers of ${\cal F}$ \cite{kruglov17,GHL_2019}; in other situations, there may be dependence on even powers of ${\cal G}$ (even powers avoid charge-parity (CP) symmetry violation) \cite{kruglov15,Gaete,Gaete2,Haas}. In our present contribution, we start off with a general Lagrangian that is a function of these two invariants to obtain the corresponding field equations and the energy-momentum tensor. Next, we expand the field-strength tensor around an electromagnetic background, initially considered non-uniform and time-dependent. We keep the terms of the expansion in the non-linear Lagrangian up to second order in the propagating excitation, and the corresponding field equations in presence of an external electromagnetic field are written down. The expansion displays coefficients that depend on the external background fields. The components of the energy-momentum tensor are calculated in the case of general space-time-dependent backgrounds. In the absence of external sources, plane wave solutions are used to calculate the allowed frequencies, the dispersion relations and the refraction index of the photon in a uniform magnetic background. The dispersion relations, and consequently, the refraction indices both depend on the relative direction of the wave vector with respect to the external magnetic field.
%

%with the direction of the external magnetic field and the direction of the photon motion.
The group velocity of the electromagnetic wave is determined from the solutions of the frequency dictated by the dispersion relations . The variation of the photon wavelength in the Compton effect is investigated in terms of the new dispersion relations and depends, of course, on the magnetic background field. We apply these results in some particular cases of non-linear electrodynamics : Hoffmann-Infeld, generalized Born-Infeld, Logarithm electrodynamics and the Euler-Heisenberg effective Lagrangian in the regime of weak electromagnetic fields.
%

%
%\textcolor[rgb]{1,0,0}{\textcolor[rgb]{1,0,0}{
More recently, the use of astrophysical sources has shown to be a very fruitful procedure to constrain modified dispersion relations (MDRs) for photons propagating in the vacuum \cite{Bolmont,SorokinPRD}. We should also recall that, early in 2019, the Major Atmospheric Gamma Imaging Cherenkov (MAGIC) telescopes identified the GRB190114C above $0.2$ TeV. This corresponds to photons with the highest frequencies detected so far in Gamma-Ray Bursts. Photons at this energy scale may well probe the quantum structure of the vacuum, so that non-linear electrodynamic effects should be taken into account. These observations motivate the growing of interest in the activity of photonic MDRs. Non-linearity in association with the presence of strong background magnetic fields yields a rich class of MDRs which may, in turn, unveil effects of new physics beyond the Standard Model (SM) of Particles and Fundamental Interactions \cite{Acciari}.
%}
%}
%
%\textcolor[rgb]{0,0,1}{
Moreover, let us recall that, in the SM, parity violation is verified in the weak-interaction sector. However, physics beyond the SM may well be sensitive to parity transformations. In this context, we could seek for evidence of parity-violating physics by inspecting MDRs in a special class on non-linear extensions of electrodynamic models, namely, the ones which explicitly depend on the special Lorentz- and gauge-invariant quantity ${\cal G}$ which, appearing with an odd power in a Lagrangian density, signals parity-symmetry breaking. This directly addresses us to the set of Planck 2018 Polarization Data, which could become a rich laboratory to constrain parity-violating non-linear extensions of electromagnetism, which may, in turn, be opening up trends to see for new physics beyond the SM \cite{Minami}.
%}
%

%\bibitem{Garcia} Eloy Ayón-Beato and Alberto García {\it The Bardeen Model as a Nonlinear Magnetic Monopole}, Phys. Lett. B {\bf 493} (2000) 149-152 [arXiv:gr-qc/0009077v1].

%\bibitem{Baier} R. Baier and P. Breintenlohner, {\it The Vacuum Refraction Index in the Presence of External Fields}, Nuovo Cimento Vol. XLVII B, 17. 1 (1967).

%
We organize our paper as follows. In Section \ref{sec2}, we give highlights of the non-linear electrodynamics framework, the corresponding field equations and energy-momentum tensor in the presence of general electric and magnetic background fields.
%In section \ref{sec3},the energy-momentum tensor in the presence of a general background is obtained.
Section \ref{sec3} focus on the frequencies for the plane wave solutions, the photonic dispersion relations in presence of a uniform magnetic field and the consequences of this background on the kinematical description of the Compton effect. In Section \ref{sec4}, we apply the results of the previous Section for a special non-linear ED depending only on the ${\cal F}$-invariant. In Section \ref{sec5}, we discuss the results by exploiting other cases of non-linear ED known in the literature that depend on both the ${\cal F}$- and ${\cal G}$-invariants.
Our Conclusions and Final Remarks are cast in Section \ref{sec6}.

The convention for the metric we adopt is $\eta^{\mu\nu}=\mbox{diag}\left(+1,-1,-1,-1\right)$. We choose to work with natural units: $\hbar=c=1$ and $4\pi\epsilon_0=1$. In this unit system, the electric and magnetic fields have squared-energy dimension. The conversion of Volt/m and Tesla (T) to the natural system is as follows:  $1 \, \mbox{Volt/m}=2.27 \times 10^{-24} \, \mbox{GeV}^2$ and $1 \, \mbox{T} =  6.8 \times 10^{-16} \, \mbox{GeV}^2$, respectively.

%%%%%%%%%%%%%%%%%%%%%%%%%%%%%%%%%%%%%%%%%%%%%%%%%%%%%%%%%%%%%%%%%%%%%%%%%%%%%%%%%%%%%%%%%%%%%%%%%%%%%%%%%%%%%%%%%%%%%%%%%%%%%%%%%%%%%%%%%%%%%%%%%%%%%%%%%%%%%%%%%%%%%%%%%%%%%%%%%%%%%%%%%%%%%%%%%%%%%%%%%%%%%%%%%%%%%%%%%%%%%%%%%%%%%%%%%%%%%%%%%%%%%%%%%%%%%%%%%%%%%%%%%%%%%%%%%%%%%%%%%%%%%%%%%%%%%%%%%%%%%%%%%%%%%%%%%%%%%%%%%%%%%%%%%%%%%%%%%%%%%%%%%%%%%%%%%%%%%%%%%%%%%%%%%%%%%%%%%%%%%%%%%%%%%%%%%%%%%%%%%%%%%%%%%%%%%%%%%%%%%%%%%%%%%%%%%%%%%%%%%%%%%%%%%%%%%%%%%%%%%%%%%%%%%%%%%%%%%%%%%%%%%%%%%%%%%%%%%%%%%%%%%%%%%%%%%%%%%%%%%%%%%%%%%%%%%%%%%%%%%%%%%%%%%%%%%%%%%%%%%%%%%%%%%%%%%%%%%%%%%%%%%%%%%%%%%%%%%%%%%%%%%%%%%%%%%%%%%%%%%%%%%%%%%%%%%%%%%%%%%%%%%%%%%%%%%%%%%%%%%%%%%%%%%%%%%%%%%%%%%%%%%%%%%%%%%%%%%%%%%%%%%%%%%%%%%%%%%%%%%%%%%%%%%%%%%%%%%%%%%%%%%%%%

%
\section{A quick glance at a general non-linear electrodynamic model}
\label{sec2}
We start off the description of the non-linear electrodynamics through the most general Lagrangian,
$\mathcal{L}$, written as a function of the Lorentz- and gauge-invariant bilinears, ${\cal F}$ and ${\cal G}$,
defined, respectively, as follows below \cite{Plebanski,Boillat,Birula1,Birula2} :
\begin{subequations}
\begin{eqnarray}
{\cal F}\!&=&\!-\frac{1}{4} \, F_{\mu\nu}^{2}=\frac{1}{2} \, \left( \, {\bf E}_{0}^2-{\bf B}_{0}^2 \, \right) \; ,
\label{invF}
\\
%\hspace{0.2cm} \mbox{and} \hspace{0.2cm}
{\cal G}\!&=&\!-\frac{1}{4} \, F_{\mu\nu}\widetilde{F}^{\mu\nu}={\bf E}_{0}\cdot{\bf B}_{0} \; ,
\label{invG}
\end{eqnarray}
\end{subequations}
where $F^{\mu\nu}=\partial^{\mu}A^{\nu}-\partial^{\nu}A^{\mu}=\left( \, -E_{0}^{\,\,i} \, , \, -\epsilon^{ijk} B_{0}^{\,\,k} \, \right)$
is the skew-symmetric field-strength tensor, $\widetilde{F}^{\mu\nu}=\epsilon^{\mu\nu\alpha\beta}F_{\alpha\beta}/2=\left( \, -B_{0}^{\,\,i} \, , \, \epsilon^{ijk} E_{0}^{\,\,k} \, \right)$
its corresponding dual tensor, which satisfies the Bianchi identity $\partial_{\mu}\widetilde{F}^{\mu\nu}=0$.
We decompose the $A^{\mu}$ potential as $A^{\mu}=a^{\mu}+A_{B}^{\;\;\,\,\mu}$, where $a^{\mu}$ is identified as
the photon field, and $A_{B}^{\;\;\,\,\mu}$ is a background potential. As consequence of this decomposition, the tensor $F^{\mu\nu}$ is written as $F^{\mu\nu}=f^{\mu\nu}+F_{B}^{\;\;\,\mu\nu}$, in which $f^{\mu\nu}=\partial^{\mu}a^{\nu}-\partial^{\nu}a^{\mu}=\left( \, -e^{i} \, , \, -\epsilon^{ijk} \, b^{k} \, \right)$ is the electromagnetic field-strength tensor of the propagating excitation, whereas $F_{B}^{\;\,\mu\nu}=\partial^{\mu}A_{B}^{\;\,\nu}-\partial^{\nu}A_{B}^{\;\,\mu} =\left( \, -E^{i} \, , \, -\epsilon^{ijk} \, B^{k} \, \right)$ corresponds to the field-strength associated with the electric and magnetic background fields. These fields, in general, depend on the space-time coordinates. Therefore, by expanding the Lagrangian ${\cal L}\left({\cal F},{\cal G}\right)$ around the background fields and keeping terms up to the second-order in the propagating field, we get
%in the perturbation theory to yield the expression
%
\begin{eqnarray}\label{L4}
{\cal L}^{(2)} \!&=&\! -\frac{1}{4} \, c_{1} \, f_{\mu\nu}^{\, 2}
-\frac{1}{4} \, c_{2} \, f_{\mu\nu}\widetilde{f}^{\mu\nu}
%\nonumber \\
%&&
%\hspace{-0.5cm}
-\frac{1}{2} \, f_{\mu\nu}G_{B}^{\; \; \; \mu\nu}
\nonumber \\
&&
\hspace{-0.5cm}
%-\frac{1}{2} \, f_{\mu\nu}G_{B}^{\; \; \; \mu\nu}
+\frac{1}{8} \, Q_{B}^{\; \; \; \mu\nu\kappa\lambda}f_{\mu\nu}f_{\kappa\lambda}
%+\frac{1}{8} \, K_{B}^{\; \; \; \mu\nu\kappa\lambda}f_{\mu\nu}f_{\kappa\lambda}
%\nonumber \\
%&&
%\hspace{-0.7cm}
%+\frac{1}{8} \, T_{B}^{\; \; \; \mu\nu\kappa\lambda}f_{\mu\nu}\widetilde{f}_{\kappa\lambda}
%\nonumber \\
%&&
%\hspace{-0.7cm}
%+\frac{1}{8} \, R_{B}^{ \; \; \; \mu\nu\kappa\lambda\rho\sigma}f_{\mu\nu}f_{\kappa\lambda}f_{\rho\sigma}
%\nonumber \\
%&&
%\hspace{-0.7cm}
%+\frac{1}{16} \, S_{B}^{ \; \; \; \mu\nu\kappa\lambda\rho\sigma\omega\tau}f_{\mu\nu}f_{\kappa\lambda}f_{\rho\sigma}f_{\omega\tau}
-J_{\mu} \, a^{\mu} - J_{\mu} \, A_{B}^{\;\;\,\,\mu} \; ,
\end{eqnarray}
where the background tensors are defined by
\begin{eqnarray}
G_{B}^{\; \; \; \mu\nu} \!\!&=&\!\! c_{1} \, F_{B}^{\, \; \; \mu\nu}+c_{2} \, \widetilde{F}_{B}^{\, \; \; \mu\nu} \; ,
\nonumber \\
%K_{B}^{\; \; \; \mu\nu\kappa\lambda} \!\!&=&\!\! d_{1} \, F_{B}^{\; \; \, \mu\nu}F_{B}^{\; \; \, \kappa\lambda}
%+d_{2} \, \widetilde{F}_{B}^{\, \; \; \mu\nu}\widetilde{F}_{B}^{\, \; \; \kappa\lambda} \; ,
%\nonumber \\
%T_{B}^{ \; \; \; \mu\nu\kappa\lambda} \!\!&=&\!\! d_{3} \, F_{B}^{\, \; \; \mu\nu}F_{B}^{\, \; \; \kappa\lambda} \; ,
%\nonumber \\
Q_{B}^{\;\;\,\, \mu\nu\kappa\lambda} \!\!&=&\!\! d_{1} \, F_{B}^{\; \; \, \mu\nu}F_{B}^{\; \; \, \kappa\lambda}
+d_{2} \, \widetilde{F}_{B}^{\, \; \; \mu\nu}\widetilde{F}_{B}^{\, \; \; \kappa\lambda}+
\nonumber \\
&&
\hspace{-0.8cm}
+d_{3} \, F_{B}^{\, \; \; \mu\nu}\widetilde{F}_{B}^{\, \; \; \kappa\lambda}
+d_{3} \, \widetilde{F}_{B}^{\, \; \; \mu\nu}F_{B}^{\, \; \; \kappa\lambda} \; ,
\hspace{0.5cm}
\end{eqnarray}
and $J^{\mu}$ is a classical source.
By construction, $G_{B}^{\; \; \; \mu\nu}=-G_{B}^{\; \; \; \nu\mu} $ and the tensor $Q_{B}^{\; \; \mu\nu\kappa\lambda}$
is antisymmetric under the exchange $\mu \leftrightarrow \nu$ or $\kappa \leftrightarrow \lambda$,
and symmetric in $\mu\nu \leftrightarrow \kappa\lambda$. The coefficients $c_{1}$, $c_{2}$, $d_{1}$, $d_{2}$ and $d_{3}$
%$M_{i} \, \left(i=1,2,3,4\right)$ and $N_{i} \, \left(i=1,2,3,4,5\right)$
are evaluated at the background fields ${\bf E}$ , ${\bf B}$ :
\begin{eqnarray}\label{coefficients}
c_{1} \!\!&=&\!\! \left.\frac{\partial{\cal L}}{\partial{\cal F}}\right|_{{\bf E},{\bf B}}
\; , \;
\left. c_{2}=\frac{\partial{\cal L}}{\partial{\cal G}}\right|_{{\bf E},{\bf B}}
\; , \;
\nonumber \\
d_{1} \!\!&=&\!\! \left.\frac{\partial^2{\cal L}}{\partial{\cal F}^2}\right|_{{\bf E},{\bf B}}
\, , \,
\left. d_{2}=\frac{\partial^2{\cal L}}{\partial{\cal G}^2}\right|_{{\bf E},{\bf B}}
\, , \,
\left. d_{3}=\frac{\partial^2{\cal L}}{\partial{\cal F}\partial{\cal G}}\right|_{{\bf E},{\bf B}}
\; , \hspace{0.3cm}
\end{eqnarray}
that, in a general situation, are space-time-dependent.
%
%**** Seção 3 aqui *****
%
%
%********
%
%\section{ The energy and momentum in the background electromagnetic field }
%\label{sec3}
%

%
%In this section,
%
Using the second-order expanded Lagrangian (\ref{L4}), we give below the energy-momentum, up to the second order in the photon field strength,
$f_{\mu\nu}$, for a general space-time-dependent background, $F_{B\mu\nu}$.
Let us consider again the Lagrangian (\ref{L4})
%
%\begin{eqnarray}
%{\cal L}^{(4)} \!\!&=&\!\! -\frac{1}{4} \, c_{1} \, f_{\mu\nu}^{\, 2}
%-\frac{1}{4} \, c_{2} \, f_{\mu\nu}\widetilde{f}^{\mu\nu}
%\nonumber \\
%&&
%\hspace{-0.5cm}
%-\frac{1}{2} \, f_{\mu\nu}G_{B}^{\, \; \;\mu\nu}
%\left( \, c_{1} F_{B}^{\, \; \;\mu\nu}+c_{2} \widetilde{F}_{B}^{\, \; \;\mu\nu} \, \right)
%\nonumber \\
%&&
%\hspace{-0.7cm}
%+\frac{1}{8} \, t_{B}^{\mu\nu\kappa\lambda}f_{\mu\nu}\widetilde{f}_{\kappa\lambda}
%+\frac{1}{8} \, \hat{Q}_{B}^{\; \; \; \mu\nu\kappa\lambda}f_{\mu\nu}f_{\kappa\lambda}
%+\frac{1}{8} \, \hat{R}_{B}^{\; \; \; \mu\nu\kappa\lambda\rho\sigma}f_{\mu\nu}f_{\kappa\lambda}f_{\rho\sigma}
%\nonumber \\
%&&
%\hspace{-0.7cm}
%+\frac{1}{16} \, \hat{S}_{B}^{\; \; \; \mu\nu\kappa\lambda\rho\sigma\omega\tau} f_{\mu\nu}f_{\kappa\lambda}f_{\rho\sigma}f_{\omega\tau}
%-J_{\mu} \, a^{\mu} \; .
%\end{eqnarray}
%
whose corresponding field equations are as follows:
%
%{\color{red}
\begin{equation}\label{EqGJ}
\partial_{\mu}\left[ c_{1} \, f^{\mu\nu}+c_{2} \, \widetilde{f}^{\mu\nu}
- \frac{1}{2} \, Q_{B}^{ \; \; \; \mu\nu\kappa\lambda}f_{\kappa\lambda} \right]=-\partial_{\mu}G_{B}^{\; \; \, \mu\nu}+J^{\nu} \, .
\end{equation}
%}
%
%where $G^{\mu\nu}$ is defined by
%
%\begin{eqnarray}
%G^{\mu\nu} \!\!&=&\!\! c_{1} \, f^{\mu\nu}+c_{2} \, \widetilde{f}^{\mu\nu}
%- \frac{1}{2} \, Q_{B}^{ \; \; \; \mu\nu\kappa\lambda}f_{\kappa\lambda} \; .
%\nonumber \\
%&&
%\hspace{-0.8cm}
%-\frac{3}{4} \, \hat{R}_{B}^{ \; \; \; \mu\nu\kappa\lambda\rho\sigma}f_{\kappa\lambda}f_{\rho\sigma}
%-\frac{1}{2} \, \hat{S}_{B}^{ \; \; \; \mu\nu\kappa\lambda\rho\sigma\omega\tau}f_{\kappa\lambda}f_{\rho\sigma}f_{\omega\tau} \; .
%\hspace{0.7cm}
%\end{eqnarray}
%
%$J_{B}^{\mu\nu}=c_{1} \, F_{B}^{\; \; \, \mu\nu}+c_{2} \, \widetilde{F}_{B}^{\; \; \, \mu\nu}$.
%
The dual tensor $\widetilde{f}^{\mu\nu}$ satisfies the Bianchi identity : $\partial_{\mu}\widetilde{f}^{\mu\nu}=0$.
We contract the field equations with $f_{\nu\alpha}$, and using the Bianchi identity for $f_{\nu\alpha}$, we obtain
the continuity equation
\begin{eqnarray}
\partial_{\mu}\Theta_{ph}^{\; \; \; \; \mu\alpha}=h^{\alpha} \; ,
\end{eqnarray}
where the energy-momentum tensor of the photon field is given by
\begin{eqnarray}\label{theta}
\Theta_{ph}^{\; \; \; \; \mu\alpha} \!&=&\! c_{1} \, f^{\mu\nu} f_{\nu}^{\;\;\alpha}
-\frac{1}{2} \, Q_{B}^{\; \; \, \, \mu\nu\kappa\lambda} f_{\kappa\lambda}f_{\nu}^{\;\;\alpha}
\nonumber \\
&&
\hspace{-0.8cm}
%-\frac{3}{4} \, \hat{R}_{B}^{\; \; \, \, \mu\nu\kappa\lambda\rho\sigma}f_{\kappa\lambda}f_{\nu}^{\;\;\alpha}f_{\rho\sigma}
%\nonumber \\
%&&
%\hspace{-0.8cm}
%-\frac{1}{2} \, \hat{S}_{B}^{\; \; \, \, \mu\nu\kappa\lambda\rho\sigma\omega\tau}f_{\kappa\lambda}f_{\nu}^{\;\;\alpha}
%f_{\rho\sigma}f_{\omega\tau}
+ \, \eta^{\mu\alpha} \, \left( \, \frac{1}{4} \, c_1 \, f_{\rho\sigma}^2
-\frac{1}{8} \, Q_{B}^{ \; \; \, \, \rho\sigma\omega\tau} f_{\rho\sigma}f_{\omega\tau} \, \right)
\; , \; \;
\end{eqnarray}
and the vector $h^{\alpha}$  is
\begin{eqnarray}\label{JB}
h^{\alpha} \!\!&=&\!\! J_{\nu}f^{\nu\alpha}-\left( \partial^{\mu}G_{B\mu\nu} \right)f^{\nu\alpha}
+\frac{1}{4}\left(\partial^{\alpha}c_{1}\right)f_{\mu\nu}^{\, 2}
\nonumber \\
&&
\hspace{-0.5cm}
+\frac{1}{4}\left(\partial^{\alpha}c_{2}\right)\widetilde{f}_{\mu\nu}f^{\mu\nu}
-\frac{1}{8} \, \left(\partial^{\alpha}Q_{B}^{\; \; \, \, \mu\nu\kappa\lambda}\right)f_{\mu\nu}f_{\kappa\lambda}
%\nonumber \\
%&&
%\hspace{-0.5cm}
%-\frac{1}{8} \, \left(\partial^{\alpha}\hat{R}_{B}^{\; \; \, \, \mu\nu\kappa\lambda\rho\sigma}\right)f_{\mu\nu}f_{\kappa\lambda}f_{\rho\sigma}
%\nonumber \\
%&&
%\hspace{-0.5cm}
%-\frac{1}{8} \, \left(\partial^{\alpha}\hat{S}_{B}^{\; \; \, \, \mu\nu\kappa\lambda\rho\sigma\omega\tau}\right)f_{\mu\nu}f_{\kappa\lambda}f_{\rho\sigma}
%f_{\omega\tau}
\; .
\end{eqnarray}
Notice that the topological term, the one in $c_{2}$, is cancelled in the expression for (\ref{theta}); as expected, it does not contribute  to the stress-tensor by virtue of its topological nature. In the general case, the background fields are non-homogeneous over space and time-dependent. Whenever $J^{\nu}=0$, the term of $h^{\alpha}$ is not zero and, as consequence, the components of the energy-momentum tensor are not conserved if the background fields are neither uniform nor constant in time. If we consider the background fields to be
constant and uniform, $h^{\alpha}$ is vanishing and, in this case, the energy-momentum tensor (\ref{theta}) satisfies
a continuity equation with the conserved energy density given in what follows below:
%
%In this point, we consider the case in which the electric
%background field is zero, and the magnetic one is uniform and constant given by ${\bf B}$. Under this condition, all the terms that contain
%derivatives are null in (\ref{JB}). Furthermore, if $J^{\nu}=0$, the conserved components of (\ref{theta}) are read below :
%
%\begin{subequations}
\begin{eqnarray}
\Theta_{ph}^{\; \; \; \; 00} \!\!&=&\!\! \frac{1}{2} \, c_{1} \left( {\bf e}^2+{\bf b}^2 \right)
+ \frac{1}{2} \, d_1 \, \left( {\bf e}\cdot{\bf E} \right)^2
+ \frac{1}{2} \, d_2 \, \left( {\bf e}\cdot{\bf B} \right)^2
\nonumber \\
&&
\hspace{-0.8cm}
-\frac{1}{2} \, d_{1} \left( {\bf b}\cdot{\bf B} \right)^2
-\frac{1}{2} \, d_2 \left( {\bf b}\cdot{\bf E} \right)^2
+d_3 \left( {\bf e}\cdot{\bf E} \right) \left({\bf e}\cdot{\bf B}\right)
\nonumber \\
&&
\hspace{-0.8cm}
+d_{3} \left( {\bf b}\cdot{\bf E} \right)\left( {\bf b}\cdot{\bf B} \right) \; ,
\label{theta00}
%\\
%\Theta_{ph}^{\; \; \; \; 0i} \!\!&=&\!\! c_{1} \left( {\bf e}\times{\bf b} \right)^{i}
%-d_1 \, \left( {\bf e} \cdot {\bf E} \right) \left( {\bf E}\times{\bf b} \right)^{i}
%\nonumber \\
%&&
%\hspace{-0.8cm}
%-d_1 \, \left( {\bf b} \cdot {\bf B} \right) \left( {\bf E}\times{\bf b} \right)^{i}
%%-d_2 \, \left( {\bf e} \cdot {\bf B} \right) \left( {\bf B}\times{\bf b} \right)^{i}
%\nonumber \\
%&&
%\hspace{-0.8cm}
%+d_2 \, \left( {\bf b} \cdot {\bf E} \right) \left( {\bf B}\times{\bf b} \right)^{i}
%-d_3 \, \left( {\bf e} \cdot {\bf B} \right) \left( {\bf E}\times{\bf b} \right)^{i}
%\nonumber \\
%&&
%\hspace{-0.8cm}
%-d_3 \, \left( {\bf e} \cdot {\bf E} \right) \left( {\bf B}\times{\bf b} \right)^{i}
%-d_3 \, \left( {\bf b} \cdot {\bf B} \right) \left( {\bf B}\times{\bf b} \right)^{i}
%\nonumber \\
%&&
%\hspace{-0.8cm}
%+d_3 \, \left( {\bf b} \cdot {\bf E} \right) \left( {\bf E}\times{\bf b} \right)^{i}
%\; ,
%\label{theta0i}
\end{eqnarray}
%\end{subequations}
%
where all the coefficients depend on the external fields, ${\bf E}$ and ${\bf B}$. We recover the Maxwell limit by
turning off the background fields and by taking $c_1=1$.
%$|{\bf B}| \rightarrow 0$.
%

%Notice that the energy density may become negative depending on the magnitude of the background fields
%and on the coefficients $c_1$ and $d_{i} \, (i=1,2,3)$.
The energy density can be written as
\begin{eqnarray}\label{theta00K}
\Theta_{ph}^{\; \; \; \; 00} =
\frac{1}{2} \, K_{ij} \, e_{i} \, e_{j} \, + \,
\frac{1}{2} \, \Lambda_{ij} \, b_{i} \, b_{j}   \; ,
\end{eqnarray}
where $K_{ij}$ and $\Lambda_{ij}$ are, respectively, defined by
\begin{eqnarray}\label{Kij}
K_{ij} \!\!&=&\!\! c_{1} \, \delta_{ij}
+ \, d_1 \, E_i \, E_j
+ d_2 \, B_i \, B_j +
\nonumber \\
&&
\hspace{-0.5cm}
+ \, d_3 \left( E_{i} \, B_{j} + E_{j} \, B_{i} \right) \; ,
\nonumber \\
\Lambda_{ij} \!\!&=&\!\!  c_{1} \, \delta_{ij}
- d_1 \, B_{i} \, B_{j} - d_2 \, E_{i} \, E_{j} +
\nonumber \\
&&
\hspace{-0.5cm}
+ \, d_3 \left( E_{i} \, B_{j} + E_{j} \, B_{i} \right) \; .
\end{eqnarray}
The energy density (\ref{theta00K}) is positive-definite whenever the eigenvalues of the symmetric matrices
$K_{ij}$ and $\Lambda_{ij}$ are non-negative. Let us contemplate the case $d_3=0$ and assume a purely magnetic background,
{\it i. e.}, $E_{i}=0$. These conditions shall actually be the ones we are going to work with in the forthcoming Sections. Therefore, with these assumptions, the eigenvalues of $K_{ij}$ are $c_1$, $c_1+\left(d_2-|d_2|\right){\bf B}^2/2$ and
$c_1+\left(d_2+|d_2|\right){\bf B}^2/2$, and the eigenvalues of $\Lambda_{ij}$ are
$c_1$, $c_1$ and $c_1-d_1\,{\bf B}^2$, respectively. If $d_2<0$ or $d_2>0$, to ensure positive eigenvalues, the following conditions
should be fulfilled:
\begin{eqnarray}\label{condEp}
c_{1}>0
\; \; , \; \;
c_1 - d_1 \, {\bf B}^2>0
\;\; \mbox{and} \;\;
c_1+d_2 \, {\bf B}^2>0 \; .
\end{eqnarray}
%
%Otherwise, the energy density is negative.

To elaborate more on the positivity of the energy density, let us recall that a general non-linear Lagrangian can be expanded as an asymptotic series in powers of ${\cal F}$ and {\cal G} :
\begin{eqnarray}
{\cal L}=a_{ij} \, {\cal F}^{i} \, {\cal G}^{j} \; , \; i,j = 0, 1, 2, ... \; .
\end{eqnarray}
MED corresponds to $i=1, j=0$.  However, except for $a_{10}=1$, all the coefficients of the expansion above are small, for they describe  tiny non-linear effects, even for strong external fields, such as, for example, magnetic fields in the neighborhood of magnetized astrophysical objects. So, in the energy density (\ref{theta00}) and in (\ref{Kij}), $c_1 = 1 + \delta_1$, with  $\delta_1 \ll 1$, since the latter stems from the coefficients $a_{ij}$ that extend the Maxwellian version. We are then arguing that, in (\ref{theta00}) and (\ref{Kij}), $\delta_1$ and the coefficients $d_i$ correspond all to tiny corrections, so that it is expected that the eigenvalues above are, in a wide range of situations, but not generally, are all positive. In these cases, the energy density (\ref{theta00}) is consequently non-negative, once the contribution given by the Maxwellian term, ${\bf e}^2 + {\bf b}^2$, dominates over the other terms. Nevertheless, we shall consider, further on, specific cases of non-linear models and the argument above may not work if the external fields become stronger than some critical value. In all situations, we are going to point out critical values of the external magnetic fields above which the average energy density of plane waves become negative, which is, to our sense, void of physical meaning. Therefore, for all models we shall discuss, we will be bound to consider external fields below the critical values we are going to derive, so as to undertake that the average energy density of the radiation be positive.

After the previous considerations, in the incoming Sections IV and V, we are going to apply these results and conditions
to the particular cases of Hoffman-Infeld, the generalized Born-Infeld, Logarithm and Euler-Heisenberg electrodynamics.

%
%\begin{eqnarray}
%% \nonumber to remove numbering (before each equation)
%   &=& 
%\end{eqnarray}

%The terms depending on $c_{2}$, $M_{2}$ and $M_{3}$ violate the CP symmetry in the energy density (\ref{theta00}).
%Considering just the magnetic background, these coefficients are nulls when applied
%in the non-linear models known in the literature, such as Born-Infeld, Logarithm electrodynamics and Euler-Heinsenberg.

%
\section{Photon dispersion relations and the kinematics of the Compton effect}
\label{sec3}

In this Section, we consider the expansion, up to second order in $f^{\mu\nu}$, to compute the dispersion relations and the propagation of photons in presence of external electromagnetic fields.
%To this order, the field equations for  $f^{\mu\nu}$ coupled to an external current (\ref{EqGJ}) read as follows.
%
%The action principle yields the field equation
%(\ref{EqGJ}), but now we write
%
%\begin{eqnarray}
%\partial_{\mu}G^{\mu\nu}=-\partial_{\mu}G_{B}^{\; \; \; \mu\nu}+J^{\nu} \; ,
%\end{eqnarray}
%
%where
%$G_{B}^{\; \; \; \mu\nu}:=c_{1} \, F_{B}^{\, \; \; \mu\nu}+c_{2} \, \widetilde{F}_{B}^{\, \; \; \mu\nu}$,
%and
%the
%tensor $G^{\mu\nu}$ as the combination
%
%\begin{eqnarray}
%G^{\mu\nu}&=&c_{1} \, f^{\mu\nu}+c_{2} \, \widetilde{f}^{\mu\nu}
%-\frac{1}{2} \, V_{B \; \; \; \; \kappa\lambda}^{\;\; \mu\nu} \, f^{\kappa\lambda} \; ,
%\end{eqnarray}
%
%and we have defined
%
%\begin{equation}
%V_{B \; \; \; \; \kappa\lambda}^{\;\; \mu\nu}=K_{B \; \; \; \; \kappa\lambda}^{\;\;\, \mu\nu}
%+\frac{1}{2} \, T_{B}^{\;\; \mu\nu\alpha\beta} \, \epsilon_{\alpha\beta\kappa\lambda}
%+\frac{1}{2} \, \epsilon^{\mu\nu\alpha\beta} \, T_{B\alpha\beta\kappa\lambda} \; .
%\end{equation}
%
%
Writing the components of the field-strength tensors in terms of the photon electric and magnetic fields, $\left( \, {\bf e} \, , \, {\bf b} \, \right)$, the background given by $\left( \, {\bf E} \, , \, {\bf B} \, \right)$,
and the source components $J^{\mu}=\left( \, \rho \, , \, {\bf J} \, \right)$, the field equations (\ref{EqGJ}) read as follows below:
%
%\begin{subequations}
%\begin{widetext}
%\begin{eqnarray}\label{eqEB}
%\nabla\cdot{\bf e}+{\bm \chi} \cdot \nabla\left( {\bf E}\cdot{\bf e}
%-{\bf B}\cdot{\bf b} \right)
%+
%\nonumber \\
%+\,{\bm \lambda} \cdot \nabla\left( {\bf B}\cdot{\bf e}
%+{\bf E}\cdot{\bf b} \right)=-\nabla\cdot{\bf E}+\frac{1}{c_{1}} \frac{\rho}{\varepsilon_{0}}
%\; , \;
%\nonumber \\
%\nabla\times{\bf e}+\partial_{t}{\bf b}={\bf 0} \; , \;
%\\ \label{eqEB2}
%\nabla\cdot{\bf b}=0
%\; , \;
%\nabla\times{\bf b}
%+ {\bm \xi}\times\nabla\left( {\bf E}\cdot{\bf e}-{\bf B}\cdot{\bf b}\right)
%+ {\bm \mu}\times\nabla\left( {\bf B}\cdot{\bf e}+ {\bf E}\cdot{\bf b} \right)
%\nonumber \\
%=\partial_{t}{\bf e}+
%\nonumber \\
%+{\bm \chi} \, \partial_{t}\left( {\bf E}\cdot{\bf e}-{\bf B}\cdot{\bf b} \right)
%+{\bm \lambda} \, \partial_{t}\left( {\bf B}\cdot{\bf e}+{\bf E}\cdot{\bf b} \right)
%-\nabla\times{\bf B}+\frac{\mu_{0}}{c_{1}} \, {\bf J} \; ,
%\end{eqnarray}
%\end{widetext}
%\end{subequations}
%
%{\color{red}
\begin{subequations}
%\begin{widetext}
\begin{eqnarray}
\nabla\cdot{\bf e}+\left(\frac{d_{1}}{c_1} \, {\bf E} + \frac{d_{3}}{c_{1}} \, {\bf B}\right) \cdot \nabla ({\bf E} \cdot {\bf e}-{\bf B} \cdot {\bf b})+
\nonumber \\
+\, \left( \frac{d_{2}}{c_{1}} \, {\bf B}+\frac{d_{3}}{c_{1}} \, {\bf E} \right) \cdot \nabla ({\bf B} \cdot {\bf e}+{\bf E} \cdot {\bf b})=
\nonumber \\
=-\nabla\cdot{\bf E}+\frac{\rho}{c_1}
\; , \; \;\;\;\;
\label{eqEB} \\
\nabla\times{\bf e}+\partial_{t}{\bf b}={\bf 0} \; \; , \; \;
\nabla\cdot{\bf b}=0
\; , \;
\hspace{1.0cm}
\label{eqEB2} \\
\nabla\times{\bf b}
+ \left( -\frac{d_{1}}{c_{1}} \, {\bf B} + \frac{d_{3}}{c_{1}} \, {\bf E} \right) \times\nabla ({\bf E} \cdot {\bf e}-{\bf B} \cdot {\bf b})
%+
\nonumber \\
+ \left( \frac{d_{2}}{c_{1}} \, {\bf E} - \frac{d_{3}}{c_{1}} \, {\bf B}  \right) \times\nabla ( {\bf B} \cdot {\bf e}+{\bf E} \cdot {\bf b})
=\partial_{t}{\bf e}+
\hspace{0.5cm}
\nonumber \\
+\left(\frac{d_{1}}{c_1} \, {\bf E} + \frac{d_{3}}{c_{1}} \, {\bf B}\right) \, \partial_{t}({\bf E} \cdot {\bf e}-{\bf B} \cdot {\bf b})+
\hspace{0.5cm}
\nonumber \\
+\left( \frac{d_{2}}{c_{1}} \, {\bf B}+\frac{d_{3}}{c_{1}} \, {\bf E} \right) \, \partial_{t}( {\bf B} \cdot {\bf e}+{\bf E} \cdot {\bf b})
-\nabla\times{\bf B}+ \, \frac{{\bf J}}{c_1} \; .
\hspace{0.4cm}
\label{eqEB3}
\end{eqnarray}
%\end{widetext}
\end{subequations}
%
%}
%
%where
%we have restored $c^2=(\mu_{0}\varepsilon_{0})^{-1}=3\times 10^{5} \, \mbox{km/s}$, and
%we have defined the scalar functions $F$ and $G$, and the background vectors ${\bm \chi}$, ${\bm \lambda}$, ${\bm \xi}$ and ${\bm \mu}$ as
%
%\begin{eqnarray}
%F \!&=&\! {\bf E} \cdot {\bf e}-{\bf B} \cdot {\bf b}
%\; \; , \; \;
%\nonumber \\
%G = {\bf B} \cdot {\bf e}+{\bf E} \cdot {\bf b} \; ,
%\nonumber \\
%{\bm \chi} \!&=&\! \frac{d_{1}}{c_1} \, {\bf E} + \frac{d_{3}}{c_{1}} \, {\bf B}
%\; \; , \; \;
%{\bm \lambda}=\frac{d_{2}}{c_{1}} \, {\bf B}+\frac{d_{3}}{c_{1}} \, {\bf E} \; ,
%\nonumber \\
%{\bm \xi} \!&=&\! -\frac{d_{1}}{c_{1}} \, {\bf B} + \frac{d_{3}}{c_{1}} \, {\bf E}
%\; \; , \; \;
%{\bm \mu}= \frac{d_{2}}{c_{1}} \, {\bf E} - \frac{d_{3}}{c_{1}} \, {\bf B} \; .
%\; \; \; \;
%\end{eqnarray}
%
%Notice that $F \mapsto -G$, $G \mapsto F$, ${\bm \chi} \mapsto {\bm \xi}$ and ${\bm \lambda} \mapsto {\bm \mu}$, whenever we perform
%the replacements ${\bf E} \rightarrow -{\bf B}$ and ${\bf B} \rightarrow {\bf E}$.
Since the Bianchi identity remains valid in non-linear electrodynamics, the divergent of ${\bf b}$ and the rotational of ${\bf e}$ keep
like in Maxwell electrodynamics.
The particular case of a Lagrangian that depends only on the invariant ${\cal F}$
in the presence of the background fields ${\bf E}$ and ${\bf B}$, we have $c_{1} \neq 0$, $d_{1} \neq 0$ and
$c_{2}=d_{2}=d_{3}=0$. The simplest case is Maxwell electrodynamics, where the Lagrangian is given by ${\cal F}$, the first
coefficient is $c_1=1$, and the other coefficients of the expansion are all vanishing. We shall consider in this
paper the case of a purely magnetic background, {\it i. e.}, ${\bf E}=0$. Whenever the non-linear model also exhibits dependence
on ${\cal G}$, this dependence must be quadratic (or an even power) in ${\cal G}^2$ to insure the charge-parity
symmetry. This fact happens in non-linear electrodynamics such as the generalized Born-Infeld, Logarithm and ArcSinh theories.
If there is no electric background, we obtain $c_2=0$ and the CP-symmetry is recovered.
We work with a uniform external magnetic field, ${\bf B}$, and, as consequence, the coefficients are also uniform and constant in time.
Other important fact is that whenever the electric background field is not present, $d_{3}=0$ for all the examples of non-linear electrodynamics
in the literature. Thereby, using that $d_3=0$, the equations (\ref{eqEB}) - (\ref{eqEB3}) with no classical
sources read as below :
\begin{subequations}
%\begin{widetext}
\begin{eqnarray}
\nabla\cdot{\bf e}+\, \frac{d_{2}}{c_1} \, {\bf B} \cdot \nabla\left( {\bf B}\cdot{\bf e} \right)=0 \; ,
%-\frac{d_{3}}{c_{1}} \, c\, {\bf B} \cdot \nabla\left( {\bf B}\cdot{\bf b} \right) =0 \; ,
%\hspace{0.3cm}
\label{eqB1} \\
\nabla\times{\bf e}+\partial_{t}{\bf b}={\bf 0}
\hspace{0.3cm} , \hspace{0.3cm}
\nabla\cdot{\bf b}=0 \; ,
\hspace{0.3cm}
\label{eqB2} \\
\nabla\times{\bf b}
%\nonumber \\
%+
%\nonumber \\
+\frac{d_1}{c_{1}} \, {\bf B}\times\nabla \left( {\bf B}\cdot{\bf b}\right)
%-\frac{d_{3}}{c_1} \, \frac{{\bf B}}{c} \times \nabla \left( {\bf B}\cdot{\bf e} \right)
=\partial_{t}{\bf e}
+
\nonumber \\
%-\frac{d_3}{c_1} \, \frac{{\bf B}}{c} \, \partial_{t}\left({\bf B}\cdot{\bf b} \right)
%\nonumber \\
+\frac{d_2}{c_1} \, {\bf B} \, \partial_{t}\left( {\bf B}\cdot{\bf e}\right) \; .
%-\nabla\times{\bf B} \; . (foi retirado este termo, já que B é constante)
\label{eqB3}
\hspace{0.5cm}
\end{eqnarray}
%\end{widetext}
\end{subequations}
The usual Maxwell equations are obtained for $d_{1}=d_{2}=0$ and $c_{1}=1$,
which is equivalent to taking $|{\bf B}| \rightarrow 0$ in (\ref{eqB1}) and (\ref{eqB3}).
Considering plane wave solutions, ${\bf e}({\bf x},t)={\bf e}_{0} \, e^{i \, \left({\bf k}\cdot{\bf x}-\omega t\right)}$ and ${\bf b}({\bf x},t)={\bf b}_{0} \, e^{i \, \left({\bf k}\cdot{\bf x}-\omega t\right)}$ in (\ref{eqB1}), (\ref{eqB2}) and  (\ref{eqB3}), the relation between the frequency, $\omega$, and the wave vector, ${\bf k}$, can be written in a matrix form:
\begin{eqnarray}\label{EqMatrix}
M_{ij} \, e_{0j}=0 \; ,
\end{eqnarray}
where $e_{0j} \, (j=1,2,3)$ are the components of the amplitude of the electric field, ${\bf e}_{0}$. The matrix elements $M_{ij}$ take the form
%
%{\color{red}
\begin{equation}\label{Mij}
M_{ij}=\alpha \, \delta_{ij}+u_{i} \, v_{j} + w_{i} \, B_{j} \; ,
\end{equation}
%}
%
where the coefficients $\alpha$, $u_{i}$, $v_{i}$ and $w_{i}$ are, respectively, defined by
%
%{\color{red}
\begin{eqnarray}
\alpha \!&=&\! \omega^2-{\bf k}^2
\hspace{0.2cm} , \hspace{0.2cm}
\nonumber \\
{\bf u} &=& \frac{d_1}{c_1} \, {\bf B} \times {\bf k}
%+ \frac{d_3}{c_1} \, \frac{\omega}{c} \, {\bf B}
%+\frac{d_3}{c_1} \, \left( {\bf B}\cdot{\bf k} \right) \, \frac{c\,{\bf k}}{\omega}
\; , \;
%\nonumber \\
{\bm v} = {\bf B} \times {\bf k} \; ,
\nonumber \\
{\bm w} \!&=&\! \frac{d_2}{c_1} \, \omega^2 \, {\bf B}
%-\frac{d_3}{c_1} \, \frac{\omega}{c} \left({\bf B} \times {\bf k}  \right)
-\frac{d_2}{c_1} \, \left({\bf B} \cdot {\bf k}  \right) {\bf k} \; .
%\nonumber \\
%{\bm t} \!&=&\! {\bf B} \; .
\end{eqnarray}
%
%}
%
%Note that the previous components are invariant under the changes $d_{1} \leftrightarrow d_{2}$, ${\bf B} \rightarrow {\bf E}/c$.
%However, in the expression of dispersion relation, $d_{3}$ appears squared, and so, we do not worry about $d_{3}$ appearing with $+$ or with $-$.
%
The matrix equation (\ref{EqMatrix}) has non-trivial solutions only if the $M$-matrix is singular.
It can be cast in the form
%
%{\color{red}
\begin{eqnarray}
\mbox{det}M=\alpha\left[\left(\alpha+ {\bf u}\cdot{\bf v}\right)\left(\alpha+{\bf w}\cdot{\bf B}\right)
-\left({\bf u}\cdot{\bf B} \right) \left({\bf v}\cdot{\bf w} \right) \right] \; , \; \; \;
\end{eqnarray}
%
%}
%
and the condition $\mbox{det}M=0$ leads to the usual photon dispersion relation $\omega^2=|{\bf k}|^2$ as one of the solutions. This is so by virtue of gauge invariance, which, in a particle scenario, corresponds to the presence of the genuine (zero mass) photon. Along with this possibility, there appear other solution as the zeroes of the polynomial equation that follows:
\begin{eqnarray}\label{eqomega4}
P \, \omega^4+Q \, \omega^2+R=0 \; ,
\end{eqnarray}
where
\begin{eqnarray}
P \!\!&=&\!\! 1+\frac{d_2}{c_1} \, {\bf B}^2 \; ,
\nonumber \\
Q \!\!&=&\!\! -2{\bf k}^2+\frac{d_{1}}{c_{1}} \, \left({\bf B} \times {\bf k}\right)^2
-\frac{d_{2}}{c_{1}} \, \left[{\bf B}^2 \, {\bf k}^2+\left( {\bf B}\cdot{\bf k}  \right)^2\right]
\nonumber \\
&&
\hspace{-0.5cm}
+\frac{d_1d_2}{c_1^{2}} \, {\bf B}^2 \left({\bf B} \times {\bf k}\right)^2 \; ,
\nonumber \\
R \!\!&=&\!\! {\bf k}^4-\frac{d_{1}}{c_{1}} \, {\bf k}^2\left( {\bf B}\times{\bf k}  \right)^2
+\frac{d_2}{c_1} \, {\bf k}^2 \left( {\bf B}\cdot{\bf k}  \right)^2
\nonumber \\
&&
\hspace{-0.5cm}
-\frac{d_1d_2}{c_1^{2}} \, \left({\bf B}\cdot{\bf k}\right)^2 \left({\bf B} \times {\bf k}\right)^2
\; .
\end{eqnarray}
%
%
%\begin{eqnarray}
%P \!\!&=&\!\! 1+\frac{d_1}{c_1} \, \frac{{\bf E}^2}{c^2}+\frac{d_2}{c_1} \, {\bf B}^2 \; ,
%\nonumber \\
%Q \!\!&=&\!\! -2{\bf k}^2+\frac{d_{1}-d_{2}}{c_{1}} \, \frac{{\bf E}^2}{c^2} \, {\bf k}^2
%-\frac{d_1+d_2}{c_1} \, \left( \frac{{\bf E}}{c}\cdot{\bf k} \right)^2
%\nonumber \\
%&&
%\hspace{-0.5cm}
%+\frac{d_1d_2+d_{3}^{2}}{c_1^{2}} \, \frac{{\bf E}^2}{c^2} \left( \frac{{\bf E}}{c} \times {\bf k}\right)^2
%\!\!+\frac{d_{1}-d_{2}}{c_{1}} \, {\bf B}^2 \, {\bf k}^2
%\nonumber \\
%&&
%\hspace{-0.5cm}
%-\frac{d_1+d_2}{c_1} \, \left( {\bf B}\cdot{\bf k}  \right)^2
%+\frac{d_1d_2+d_{3}^{2}}{c_1^{2}} \, {\bf B}^2 \left({\bf B} \times {\bf k}\right)^2 \; ,
%\nonumber \\
%R \!\!&=&\!\! {\bf k}^4-\frac{d_{2}}{c_{1}} \, {\bf k}^2\left( \frac{{\bf E}}{c}\cdot{\bf k}  \right)^2
%+\frac{d_1}{c_1} \, {\bf k}^2 \left( \frac{{\bf E}}{c}\cdot{\bf k}  \right)^2
%\nonumber \\
%&&
%\hspace{-0.5cm}
%+\frac{d_{3}^{2}-d_1d_2}{c_1^{2}} \, \left(\frac{{\bf E}}{c}\cdot{\bf k}\right)^2 \left( \frac{{\bf E}}{c} \times {\bf k}\right)^2
%\!\!-\frac{d_{1}}{c_{1}} \, {\bf k}^2\left( {\bf B}\times{\bf k}  \right)^2
%\nonumber \\
%&&
%\hspace{-0.5cm}
%+\frac{d_2}{c_1} \, {\bf k}^2 \left( {\bf B}\cdot{\bf k}  \right)^2
%+\frac{d_{3}^{2}-d_1d_2}{c_1^{2}} \, \left({\bf B}\cdot{\bf k}\right)^2 \left({\bf B} \times {\bf k}\right)^2
%\; .
%\end{eqnarray}
%
The roots of (\ref{eqomega4}) are $\omega_{1}^{(\pm)}=\pm \, \omega_{1}({\bf k})$ and $\omega_{2}^{(\pm)}=\pm \, \omega_{2}({\bf k})$, whose frequencies are shown below:
\begin{subequations}
\begin{eqnarray}
\omega_{1}({\bf k}) \!&=&\! |{\bf k}| \, \sqrt{1-\frac{d_1}{c_1} \, ({\bf B}\times \hat{{\bf k}})^2 }  \; ,
\label{omega1Bpm} \\
\omega_{2}({\bf k}) \!&=&\! |{\bf k}| \, \sqrt{1-\frac{d_2 \, ({\bf B}\times \hat{{\bf k}})^2 }{ c_1+d_{2}\,{\bf B}^2 } }  \; .
%\nonumber \\
%\omega_{2}^{(\pm)} \!&=&\! \pm \, c \, \sqrt{ \frac{-Q+\sqrt{Q^2-4 \, P \, R}}{2P} } \; .
\label{omega2Bpm}
\end{eqnarray}
\end{subequations}
The usual photon frequencies are recovered in the limit $d_{1} \rightarrow 0$ and $d_{2} \rightarrow 0$,
or, equivalently, whenever $|{\bf B}| \rightarrow 0$. Notice that, if the non-linear theory only depends on the
${\cal F}$-invariant, $d_2=0$ and the second solution recovers the usual dispersion relation.
The frequencies (\ref{omega1Bpm}) and (\ref{omega2Bpm}) are real if $c_1>d_1({\bf B}\times \hat{{\bf k}})^2$ and
$c_1+d_2({\bf B}\cdot\hat{{\bf k}})^2>0$, respectively. The (magnetized) vacuum refraction index is given as the
inverse of the phase velocity, $\omega_{i}/|{\bf k}| \, (i=1,2)$ :
\begin{subequations}
\begin{eqnarray}
n_{1}^{-1} \!&=&\! \sqrt{1-\frac{d_1}{c_1} \, ({\bf B}\times \hat{{\bf k}})^2 } \; ,
\label{n1} \\
n_{2}^{-1} \!&=&\! \sqrt{1-\frac{d_2 \, ({\bf B}\times \hat{{\bf k}})^2 }{ c_1+d_{2}\,{\bf B}^2 } } \; .
\label{n2}
\end{eqnarray}
\end{subequations}
%
%If we multiply the solutions (\ref{omegaBpm}) by $\hbar$, and
From the De Broglie duality correspondence, the energy-momentum relations for the propagating excitations read as below:
\begin{subequations}
\begin{eqnarray}
E_{1}^2 \!&=&\! {\bf p}^2 \left[ \, 1-\frac{d_1 }{c_1} \, ({\bf B}\times \hat{{\bf p}})^2 \, \right] \; ,
\label{E1} \\
E_{2}^2 \!&=&\! {\bf p}^2 \left[ \, 1-\frac{d_2 \, ({\bf B}\times \hat{{\bf p}})^2 }{ c_1+d_{2}\,{\bf B}^2 } \, \right] \; .
\label{E2}
\end{eqnarray}
\end{subequations}
%

%Maxwell case is recovered when $c_1=1$ and $d_{1}=d_{2}=0$ in (\ref{eqomega4}).
%The result (\ref{omega2B}) is recovered by taking $d_{2}=d_{3}=0$.
%%
%

%
%
The group velocity associated with the previous frequencies can be read off from the equations cast in what follows:
\begin{subequations}
\begin{eqnarray}
\left.
{\bf v}_{g}
\right|_{\omega=\omega_{1}}
\!&=&\! \frac{ c_1 \, \hat{{\bf k}} + d_1 \, {\bf B} \times ( {\bf B} \times \hat{{\bf k}} )}{c_{1} \, \sqrt{1-\frac{d_1}{c_1} ({\bf B}\times\hat{\bf k})^2 } }
\; ,
\\ \label{vg1}
\left.
{\bf v}_{g}
\right|_{\omega=\omega_{2}} \!&=&\! \frac{ \hat{{\bf k}} \, c_{1} + d_2 \, {\bf B} \, ({\bf B}\cdot\hat{{\bf k}}) }
{(c_1+d_2 \, {\bf B}^2)\sqrt{1-\frac{d_2({\bf B}\times\hat{\bf k})^2}{c_1+d_2{\bf B}^2} }} \; .
\label{vg2}
\end{eqnarray}
\end{subequations}
The group velocity vectors have components in the directions of $\hat{{\bf k}}$ and $\hat{{\bf B}}$.
Both results go to ${\bf v}_{g}=\hat{{\bf k}} \, \omega/|{\bf k}|$, in the limit $|{\bf B}| \rightarrow 0$,
or if we consider the Maxwellian limit. If the magnetic background, ${\bf B}$, is perpendicular to the direction of propagation, $\hat{{\bf k}}$,
the solutions also reduce to ${\bf v}_{g}=(1-d_1 \, {\bf B}^2/c_1)^{1/2}\hat{{\bf k}}$ and ${\bf v}_{g}=(1+d_2 \, {\bf B}^2/c_1)^{-1/2}\hat{{\bf k}}$, respectively.

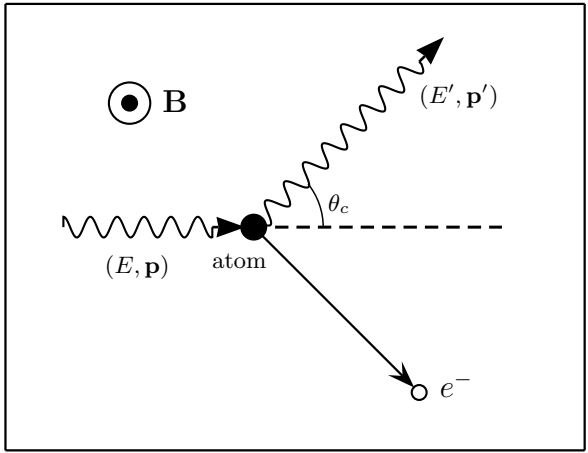
\begin{figure}[!h]%\label{figvm}
\begin{center}
\newpsobject{showgrid}{psgrid}{subgriddiv=1,griddots=10,gridlabels=6pt}
\begin{pspicture}(-3,-1.55)(4,3.95)
%\showgrid
\psset{arrowsize=0.2 2}
\psset{unit=1.1}
%
%%%%%%%%%%%%%%%%%%% Quadrado %%%%%%%%%%%%%%%%%%%%%%%%%%%%%%%
%
\psline[linecolor=black,linewidth=0.3mm](-3,3.7)(4,3.7)
\psline[linecolor=black,linewidth=0.3mm](4,3.7)(4,-1.7)
\psline[linecolor=black,linewidth=0.3mm](4,-1.7)(-3,-1.7)
\psline[linecolor=black,linewidth=0.3mm](-3,-1.7)(-3,3.7)
%
%%%%%%%%%%%%%%%%%%%%%%%%%%% Before the collision %%%%%%%%%%%%%%%%%%%%%%%%%%%%%%%%%%
%
%\pscoil[coilaspect=0,coilarm=0,coilwidth=0.25,coilheight=1.3,linecolor=black](0,0.3)(2.99,0)
%
%\rput(1.5,1.4){Photon}
%\psplot[linecolor=red,linewidth=1.0pt,plotpoints=1200]%
%{-0.065}{3.1}{x 180 mul 3.141592654 div sin x 15 mul 180 mul 3.141592654 div cos mul}
%
\psline[linecolor=black,linewidth=1.0pt,arrowsize=0.25,arrowinset=0.07]{->}(-0.5,1)(-0.1,1)
\pscoil[coilaspect=0,coilarm=0,coilwidth=0.25,coilheight=1.3,linecolor=black](-2.3,1)(-0.5,1)
\put(-1.80,0.45){$(E,{\bf p})$}
%
%\put(1.5,-0.5){$W/Z$}
%
%
\pscircle[fillstyle=solid,fillcolor=black](0,1){.15}
%
%
%%%%%%%%%%%%%%%%% After the collision %%%%%%%%%%%%%%%
%
%
\psline[linestyle=dashed,linewidth=0.4mm](0,1)(3,1)
\psline[linecolor=black,linewidth=1.0pt,arrowsize=0.25,arrowinset=0.07]{->}(2,3)(2.3,3.3)
\pscoil[coilaspect=0,coilarm=0,coilwidth=0.25,coilheight=1.3,linecolor=black](0,1)(2,3)
\pscircle[fillstyle=solid,fillcolor=white](2,-1){.09}
%
%\psline[linestyle=dashed,linewidth=0.5mm]{->}(1.5,1)(3,0)
%
\put(2,2.5){$(E^{\prime},{\bf p}^{\prime})$}
%
%\put(2.1,1.7){$F$}
%
%\put(2.1,0){$F$}
%
%%%%%%%%%%%%%%%%%%%%%%%%%%%%%%%%%%%%%%%
%
\psline[linecolor=black,linewidth=0.3mm]{->}(0,1)(1.94,-0.94)
\psarc[linewidth=0.2mm](0,1){0.84}{0}{38.5}
%
%\psline[linecolor=black,linewidth=0.3mm](3.4,2.2)(4,2.5)
%
\put(0.9,1.2){$\theta_{c}$}
%
%\psline[linecolor=black,linewidth=0.3mm](3,2)(3.75,1.63)
%\psline[linecolor=black,linewidth=0.3mm]{<-}(3.5,1.75)(4,1.5)
%
\put(-0.50,0.5){\mbox{atom}}
\put(2.25,-1.05){\large$e^{-}$}
%
%%%%%%%%%%%%%%%%%%% Magnetic field %%%%%%%%%%%%%%%%%%%%%%%%%%%%%%%%%
%
\pscircle[fillstyle=solid,fillcolor=white](-1.5,2.5){.25}
\pscircle[fillstyle=solid,fillcolor=black](-1.5,2.5){.09}
\put(-1.1,2.4){\large${\bf B}$}
%
%\psline[linecolor=black,linewidth=0.3mm]{->}(3,0)(3.75,0.37)
%\psline[linecolor=black,linewidth=0.3mm](3.4,0.2)(4,0.5)
%
%\put(3.5,0.55){$\ell_{i}^{-}$}
%
%\psline[linecolor=black,linewidth=0.3mm](3,0)(3.75,-0.37)
%\psline[linecolor=black,linewidth=0.3mm]{<-}(3.4,-0.2)(4,-0.5)
%
%\put(3.5,-0.8){$\ell_{i}^{+}$}
%
%\psline[linecolor=black,linewidth=0.3mm]{<-}(0,1)(-1.5,-0.5)
%\psline[linecolor=black,linewidth=0.3mm]{->}(1,2)(-0.4,3.4)
%
%
%Momenta
%
%\psline[linecolor=black,linewidth=0.3mm]{->}(2.2,2.7)(3.8,2.7)
%\put(3,2.6){\Large$X^{\mu}$}
%
%\psline[linecolor=black,linewidth=0.3mm]{->}(7.8,1.2)(7.8,2.8)
%\put(3,1){\Large$k$}
%
%\put(5.5,3.6){\Large$e^{-}$}
%\put(5.4,-0.1){\Large$e^{+}$}
%\put(6.7,3.1){\Large$q$}
%\put(6.7,0.7){\Large$q^{\prime}$}
%
%\put(-0.2,3.6){\Large$e^{-}$}
%\put(-0.4,-0.1){\Large$e^{+}$}
%\put(-1.1,3.1){\Large$q$}
%\put(-1.1,0.7){\Large$q^{\prime}$}
%
%\put(0.6,-0.3){$p$}
%
%\put(1.3,0){$\Gamma_{\mu}^{(3)\;abc}(p)=g_{1}f^{abc}p_{\mu}$}
%
%
\end{pspicture}
\vspace{0.2cm}
\caption{ Setup for the study of the kinematics of the Compton effect.
We consider the external magnetic field orthogonal to the
plane of the Compton array and coming out from the plane.} \label{Comptonfig}
\end{center}
\end{figure}

The Compton effect is a scattering process in which the dispersion relations (\ref{E1}) and (\ref{E2})
can be applied to study the increasing of the photon wavelength after being scattered by the electron,
taken as the target, in presence of a magnetic background. The photon initial state has wavelength $\lambda=1/|{\bf p}|$,
with energy $E$, where the relation between $E$ and the momentum, ${\bf p}$, must now follow from (\ref{E1}) or (\ref{E2}).
The physical scenario consists of a photon that propagates in the magnetic background and collides with an electron in an atom at rest. So, before the photon hits the atom, we consider that the external magnetic field affects only the photon dispersion relation. The collision causes the electron to recoil with a given energy and momentum. Then, after the scattering process takes place, the outgoing electron couples to the magnetic field and, clearly, the electron dispersion relation is no longer of a free electron. But, we are focusing on the wavelength shift and scattering angle of the photon; this is why we are not discussing the effect of the magnetic field on the dispersion relation of the emergent electron.
The setup corresponding to the Compton effect whose kinematics
we are studying here is depicted in Figure \ref{Comptonfig}.
%

%%%%%%%%%%%%%%%%% The Compton effect under the external B field %%%%%%%%%%%%%%%%%%%%%%%%%
%
After the collision, the photon trajectory is deviated by an angle $\theta_{c}$, with wavelength
$\lambda^{\prime}=1/|{\bf p}^{\prime}|$ and energy $E^{\prime}$. From the energy and linear momentum conservation,
the variation of the photon wavelength, after the collision process, turns out to be
\begin{equation}\label{deltalambda}
\lambda_{i}^{\prime}-\lambda = 2 \, \lambda_{e} \sin^2\left(\frac{\theta_c}{2}\right)+
%\nonumber \\
%&&
%\hspace{-1cm}
%+
a_{i}
%\, \frac{d_2\left|\,{\bf B} \times \hat{{\bf p}}\,\right|^2}{c_{1}+d_2\,{\bf B}^2}
\left[\lambda_{i}^{\prime}-\lambda+\frac{\lambda_{e}}{2}\frac{\left(\lambda_{i}^{\prime}-\lambda\right)^2}{\lambda_{i}^{\prime} \, \lambda} \right] \; ,
\end{equation}
where $\lambda_{e}=m_{e}^{-1}=2 \, \left(\mbox{MeV}\right)^{-1}$
%3.861 \times 10^{-13} \, \mbox{m}$
is the Compton wavelength of the electron, $a_{i} \, (i=1,2)$ means $a_1:=\left|\,{\bf B} \times \hat{{\bf p}}\,\right|^2d_1/c_1 $ and $a_2:=d_2\left|\,{\bf B} \times \hat{{\bf p}}\,\right|^2/\left(c_{1}+d_2{\bf B}^2\right)$ for both the cases of (\ref{E1}) and (\ref{E2}), respectively, and $\lambda_{i}^{\prime}$
are the wavelengths for both cases $i=1,2$ after the collision. We take the initial photon wavelength
in the range of the X-ray spectrum $\lambda= 1.52 \times \left( \, 10^{10} - 10^{13} \, \right) \, \mbox{MeV}^{-1}$.
The corresponding solutions to (\ref{deltalambda}) yielding the final wavelength of the photon are given by
\begin{widetext}
\begin{eqnarray}\label{lambdai}
\lambda_{i}^{\prime(\pm)}&=&\frac{\lambda+2\lambda_{e} \sin^2(\theta_c/2)-a_i \left( \lambda + \lambda_{e} \right)}{2-2 a_i-a_i\lambda_{e}/\lambda}
\, \pm \, \frac{\sqrt{\left(\lambda+2\lambda_{e}\sin^2(\theta_c/2)-a_i \lambda-a_i \lambda_{e}\right)^2
+ a_i \lambda_{e} \left(2\lambda-2 a_i\lambda-a_i\lambda_{e}\right)}}{2-2a_i-a_i \lambda_{e}/\lambda} \; .
\; \;
%\nonumber \\
%\lambda_{+}^{\prime}&=&\frac{\sqrt{\left(\lambda+2\lambda_{e}\sin^2(\theta/2)-a \lambda-a \lambda_{e}\right)^2
%+a\lambda_{e} \left(2\lambda-2a\lambda-a\lambda_{e}\right)}}{2-2 a-a \lambda_{e}/\lambda}
%\nonumber \\
%&&
%+\frac{\lambda+2\lambda_{e} \sin^2(\theta/2)-a\left( \lambda + \lambda_{e} \right) }{2-2 a-a\lambda_{e}/\lambda} \; .
\end{eqnarray}
\end{widetext}
%
%in which we have defined $a:=d_2\left|\,{\bf B} \times \hat{{\bf p}}\,\right|^2/\left(c_{1}+d_2{\bf B}^2\right)$ for simplify the expression.
If we assume $\lambda \gg \lambda_{e}$ in (\ref{lambdai}) , the real and positive solutions lead us to the two wavelength variations below:
\begin{subequations}
\begin{eqnarray}
\Delta\lambda_{1} \!&\simeq&\! 2 \lambda_{e} \sin^2\left(\frac{\theta_c}{2}\right) \left[\, 1-\frac{d_1}{c_1} \, |{\bf B} \times \hat{{\bf p}}|^{2} \, \right]^{-1}
%\!\!\!\!+{\cal O}\left(\frac{\lambda_{e}^2}{\lambda^2}\right)
,
\label{lambda1} \\
\Delta\lambda_{2} \!&\simeq&\! 2 \lambda_{e} \sin^2\left(\frac{\theta_c}{2}\right) \left[ \, 1-\frac{d_2 \, |{\bf B} \times \hat{{\bf p}}|^{2} }{c_1+d_2\,{\bf B}^2} \, \right]^{-1}
%\!\!\!\!+{\cal O}\left(\frac{\lambda_{e}^2}{\lambda^2}\right)
,
\label{lambda2}
\end{eqnarray}
\end{subequations}
that are positive if $c_1>d_1\,|{\bf B} \times \hat{{\bf p}}|^{2}$ and $c_1+d_2\left({\bf B} \cdot \hat{{\bf p}}\right)^2>0$.
%
%
%
%requires the conditions
%$(a_{1},a_{2}) < \sin^2(\theta_{c}/2)$ and $(a_{1},a_{2}) < \lambda/\lambda_{e}$.
%The solutions $\lambda_{i}^{\prime(-)}$ can be negative depending on some values of $\lambda$ and $a_i$, thus we can discard it.
%When $a_i \ll 1$, the positive solution can be written in terms of the variation of the photon wavelength
%
%\begin{widetext}
%\begin{eqnarray}
%\Delta\lambda_{i}
%=\lambda_{+}^{\prime}-\lambda
%\simeq 2 \lambda_{e} \, \sin^2\left(\frac{\theta_c}{2}\right)
%+
%\frac{2d_{2}\left|\,{\bf B} \times \hat{{\bf p}}\,\right|^2}{c_{1}+d_2{\bf B}^2}
%2\lambda_{e}\sin^2\left(\frac{\theta_c}{2} \right) \, \times
%\nonumber \\
%\times \,
%\frac{a_i}{\lambda} \left[ \frac{\lambda^2\cos^2(\theta_c/2)+\left( \lambda+\lambda_{e} \right)^2\sin^2(\theta_c/2)}{\lambda+ 2\lambda_{e} \sin^2(\theta_c/2) } \right] \; .
%\end{eqnarray}
%\end{widetext}
%
%\begin{eqnarray}
%\Delta\lambda_{i}
%=\lambda_{+}^{\prime}-\lambda
%\simeq 2 \lambda_{e} \, \sin^2\left(\frac{\theta_c}{2}\right)
%+
%\frac{2d_{2}\left|\,{\bf B} \times \hat{{\bf p}}\,\right|^2}{c_{1}+d_2{\bf B}^2}
%2\lambda_{e}\sin^2\left(\frac{\theta_c}{2} \right) \, \times
%\nonumber \\
%\times \,
%\frac{a_{i}}{\lambda} \left[ \frac{\lambda^2\cos^2(\theta_c/2)+\left( \lambda+\lambda_{e} \right)^2\sin^2(\theta_c/2)}{\lambda+ 2\lambda_{e} \sin^2(\theta_c/2) } \right] \; .
%\end{eqnarray}
%
Whenever $d_1,d_2 \rightarrow 0$, the standard variation of the photon wavelength in the Compton effect is recovered.
The non-linear contribution depends on the external magnetic field, ${\bf B}$, and the parameters $c_{1}$, $d_{1}$ and $d_{2}$.
These parameters, in turn, also depend on the external magnetic field and the specific dependence is governed by the non-linear electrodynamics under consideration.
\section{ Example of an ${\cal F}$-dependent electrodynamic model }
\label{sec4}
\begin{figure}[t]
%\vspace{-5pt}
\centering
\includegraphics[width=0.47\textwidth]{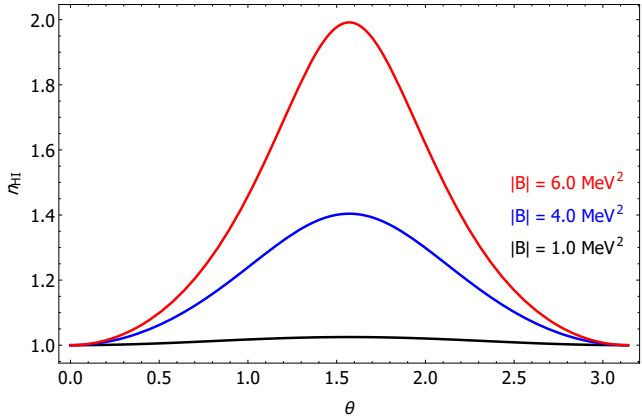}
\caption{The vacuum refraction index of the HI ED appears
as a function of the $\theta$-angle between ${\bf B}$ and the direction of $\hat{{\bf k}}$. We choose $\beta=10.0\,\mbox{MeV}^2$,
for the values $|{\bf B}|=1.0\,\mbox{MeV}^2$, $|{\bf B}|=4.0\,\mbox{MeV}^2$ and $|{\bf B}|=6.0\,\mbox{MeV}^2$.}
\label{nthetaHI}
\end{figure}

\begin{figure}[t]
%\vspace{-5pt}
\centering
\includegraphics[width=0.47\textwidth]{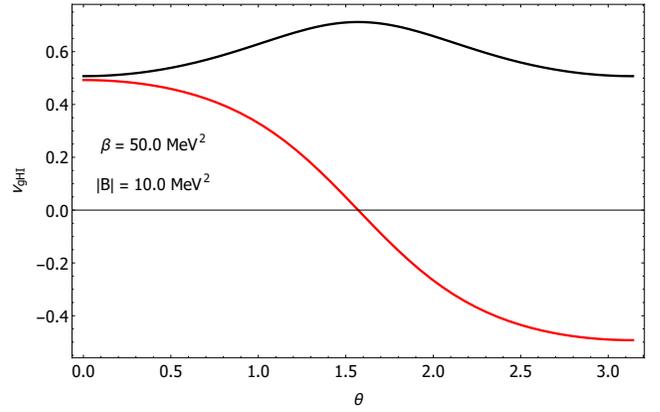}
\caption{The components of the group velocity as functions of $\theta$ in the HI ED.
The black line is the velocity component in the direction of $\hat{{\bf k}}$,
while the red line stands for the $\hat{{\bf B}}$ component.}
\label{vHI}
\end{figure}

\begin{figure}[t]
%\vspace{-5pt}
\centering
\includegraphics[width=0.49\textwidth]{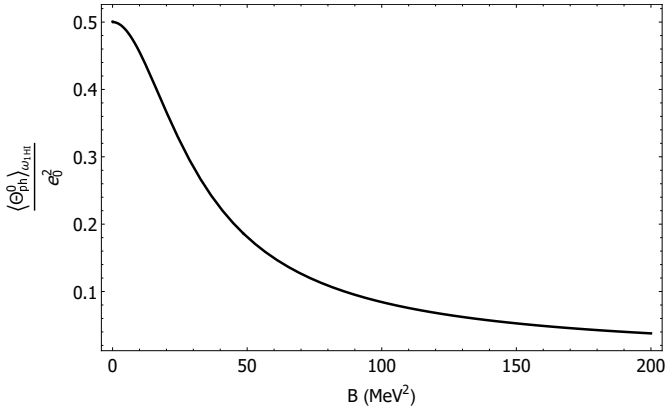}
\caption{The average (in time) of the energy density per unit of the squared amplitude of the electric field versus the external magnetic field in the HI ED.
We choose $\beta=50 \, \mbox{MeV}^2$, and the vectors $\hat{{\bf k}}$, $\hat{{\bf e}}_{0}$ and ${\bf B}$
are perpendicular to one another.}
\label{thetaHI}
\end{figure}

The Hoffmann-Infeld (HI) model is an example of non-linear electrodynamics with interesting application in the study of special black-hole
solutions \cite{Aiello}. The Lagrangian is given by
\begin{eqnarray}
{\cal L}_{HI}({\cal F})=\frac{\beta^2}{4} \, \left[ \, 1-\eta({\cal F})-\ln\eta({\cal F}) \, \right] \; ,
\end{eqnarray}
where $\eta({\cal F})$ is defined by
\begin{eqnarray}
\eta({\cal F})= \frac{4{\cal F}}{\beta^2-\beta\,\sqrt{\beta^2-8{\cal F}}} \; .
\end{eqnarray}
The parameter $\beta$ is introduced to guarantee a finite electrostatic field configuration in the case the particle-like charges.
Maxwell Electrodynamics is recovered when $\beta \rightarrow \infty$. In this Section, since the model depends
exclusively on the invariant ${\cal F}$, the coefficient vanishes, $d_2=0$. Thereby, the non-trivial dispersion relation in a magnetic
background corresponds to (\ref{E1}), which depends on $d_1$ and $c_1$. The non-trivial coefficients, $c_1$ and $d_1$, are given by
\begin{subequations}
\begin{eqnarray}
\left. c_1^{HI} \right|_{{\bf E}=0,{\bf B}} \!&=&\! \frac{\beta}{4 {\bf B}^2} \frac{ -\beta^2+2 {\bf B}^2 +\beta \, \sqrt{\beta^2+4 {\bf B}^2}}{\sqrt{\beta^2+4 {\bf B}^2}} \; ,
\hspace{0.7cm}
\label{c1HI}
\\
\left. d_1^{HI} \right|_{{\bf E}=0,{\bf B}} \!&=&\! \frac{\beta}{{\bf B}^2} \frac{2{\bf B}^2-3 \beta^2 +2\beta \sqrt{\beta^2+4 {\bf B}^2}}{\left(\beta^2+4 {\bf B}^2\right)^{3/2}}
\nonumber \\
&&
\hspace{-0.5cm}
-\frac{\beta^4}{2{\bf B}^4}\frac{\beta -\sqrt{\beta^2+4 {\bf B}^2}}{\left(\beta^2+4 {\bf B}^2\right)^{3/2}} \; .
\label{d1HI}
\end{eqnarray}
\end{subequations}
The corresponding refraction index as function of the angle $\theta$ between the magnetic background and the direction of propagation,
$\hat{{\bf k}}$, is shown in figure (\ref{nthetaHI}). We choose the following values for the magnetic background:
$|{\bf B}|=1.0\,\mbox{MeV}^2$ (black line), $|{\bf B}|=4.0\,\mbox{MeV}^2$ (blue line) and $|{\bf B}|=8.0\,\mbox{MeV}^2$ (red line).
For a strong magnetic background, {\it i. e.}, $|{\bf B}| \gg \beta$, the refraction index of the HI model is $n_{HI}=|\sec\theta|$; this refraction index vanishes if ${\bf B}$ is perpendicular to the direction $\hat{{\bf k}}$, and $n_{HI}=1$ if ${\bf B}$ is parallel to $\hat{{\bf k}}$.
The components of the group velocity (\ref{vg1}) for the HI ED as functions of $\theta$-angle are plotted in the figure (\ref{vHI}).
We choose $\beta =50.0 \mbox{MeV}^2$ and $|{\bf B}|=10.0 \mbox{MeV}^2$ in this case.
The black line stands for the component in the direction of $\hat{{\bf k}}$, whereas the red line represents
the component in the direction of magnetic field $\hat{{\bf B}}$. This component is negative for the range
$\pi/2 \leq \theta \leq \pi$. When $|{\bf B}| \gg \beta$, the group velocity has the behaviour
\begin{eqnarray}
{\bf v}_{gHI} \simeq -\frac{\beta}{|{\bf B}|} \, |\sec\theta| \, \hat{{\bf k}}+ \mbox{sgn}(\cos\theta) \, \hat{{\bf B}} \; .
\end{eqnarray}
The next analysis refers to the energy density of the HI ED in the presence of the background, ${\bf B}$. We use the result (\ref{theta00})
with the coefficients (\ref{c1HI}) and (\ref{d1HI}). Under the conditions (\ref{condEp}), the energy density in the HI ED
is positive if $|{\bf B}| < 0.972 \, \beta$, or $|{\bf B}| > \sqrt{2} \, \beta$. To illustrate the energy density (\ref{theta00}),
we consider the case in which the plane wave for the electric and magnetic fields ${\bf e}({\bf x},t)={\bf e}_{0} \, e^{i \, \left({\bf k}\cdot{\bf x}-\omega t\right)}$
and ${\bf b}({\bf x},t)={\bf b}_{0} \, e^{i \, \left({\bf k}\cdot{\bf x}-\omega t\right)}$,
respectively, propagate in a medium with an uniform and constant magnetic background.
We obtain thereby the time average of the energy density of the HI ED per unit of the squared electric field :
\begin{equation}\label{Theta00HI}
\frac{\langle \Theta_{ph}^{00} \rangle^{HI} }{{\bf e}_{0}^{2}}=\frac{1}{4} \, c_{1}^{HI}\left[1+\frac{{\bf k}^2}{\omega^2}
-\frac{d_{1}^{HI}}{c_{1}^{HI}} \frac{{\bf k}^2}{\omega^2} \left( \hat{{\bf e}}_{0}\cdot(\hat{{\bf k}} \times {\bf B}) \right)^2  \right] \; .
\; \; \;
\end{equation}
Notice that this result depends on the frequency solutions of $\omega$, that in the HI case just (\ref{omega1Bpm})
depend on the external magnetic field. We consider the vectors $\hat{{\bf k}}$, $\hat{{\bf e}}_{0}$ and ${\bf B}$
perpendicular to each other in this analysis, and we also choose $\beta=50 \mbox{MeV}^2$. Under these conditions,
the result (\ref{Theta00HI}) is plotted as function of the magnetic background in the figure (\ref{thetaHI}).
In this case, the energy density is positive for all values of the ${\bf B}$ magnitude, and goes to zero if
$|{\bf B}| \rightarrow \infty$. The limit $|{\bf B}| \rightarrow 0$ recovers the known result in Maxwell ED.
\section{Some ${\cal F}$- and ${\cal G}$-dependent electrodynamic models }
\label{sec5}
Many examples of non-linear electrodynamics with dependence on both ${\cal F}$ and ${\cal G}$
are discussed in the literature. In these cases, the coefficient $d_2 \neq 0$ and the two frequency solutions, (\ref{omega1Bpm})
and (\ref{omega2Bpm}), must be considered. The models, in general, depend on ${\cal G}^2$ (simplest case) or an even power of ${\cal G}$, so as to ensure CP-invariance.
Let us start off by contemplating the well-known case of the Euler-Heisenberg (EH) electrodynamics, described by the effective Lagrangian
\begin{eqnarray}\label{LEH}
{\cal L}_{EH}({\cal F},{\cal G})={\cal F}-
\frac{1}{8\pi^2} \int_{0}^{\infty} \frac{ds}{s^3} \, \, e^{-m^2 \, s} \, \times
\hspace{0.5cm}
\nonumber \\
\times \, \left[ (es)^2 \, {\cal G} \, \frac{\Re\cosh\left(es\sqrt{-{\cal F}+i\,{\cal G}}\right)}{\Im\cosh\left(es\sqrt{-{\cal F}+i\,{\cal G}}\right)} +\frac{2}{3} (es)^{2} {\cal F}-1 \right] \; , \; \; \;
\end{eqnarray}
where $\Re$ and $\Im$ stand for the real and imaginary parts, respectively, and $m=0.5 \, \mbox{MeV}$ is the electron mass.
In the weak field approximation, the EH Lagrangian (\ref{LEH}) is reduced to the form
\begin{eqnarray}\label{LEHapprox}
{\cal L}_{EH}({\cal F},{\cal G})\simeq{\cal F}+\frac{2\alpha^2}{45m^4} \left(\, 4 \, {\cal F}^2\,+\,7 \, {\cal G}^2 \,\right) \; ,
\end{eqnarray}
in which $\alpha=e^2=(137)^{-1}=0.00729$ is the fine structure constant. Below, we quote the coefficients $c_{1}$, $d_{1}$ and $d_{2}$ of the expansion corresponding to the truncation given by (\ref{LEHapprox}) :
\begin{eqnarray}\label{cEH}
\left. c_{1}^{EH} \right|_{{\bf E}=0,{\bf B}} \!&=&\! 1-\frac{8\alpha^2 \, {\bf B}^2}{45m^4}
\; , \;
\nonumber \\
\left.d_{1}^{EH} \right|_{{\bf E}=0,{\bf B}} \!&=&\! \frac{16\alpha^2}{45m^{4}} \; ,
\nonumber \\
\left. d_{2}^{EH} \right|_{{\bf E}=0,{\bf B}} \!&=&\! \frac{28\alpha^2}{45m^{4}}  \; .
\end{eqnarray}
Using the results (\ref{omega1Bpm}) and (\ref{omega2Bpm}) with the coefficients (\ref{cEH}),
the two solutions for the frequencies are
\begin{subequations}
\begin{eqnarray}
\omega_{1}^{EH}({\bf k}) \!\!&\simeq&\!\! |{\bf k}|\left[1- \frac{8\alpha^2}{45 m^4} ({\bf B}\times\hat{{\bf k}})^{2} \right] \; ,
\label{omegaEH1}
\\
\omega_{2}^{EH}({\bf k}) \!\!&\simeq&\!\! |{\bf k}|\left[1- \frac{14\alpha^2}{45 m^4} ({\bf B}\times\hat{{\bf k}})^{2} \right] \; ,
\label{omegaEH2}
\end{eqnarray}
\end{subequations}
and, then, the corresponding vacuum refraction index for the EH effective model, in the approximation we are working, turn out given by
\begin{subequations}
\begin{eqnarray}
n_{1}^{EH}
%\!\!\!&=&\!\!\!
%\left[ 1-\frac{16\alpha^2 \, ({\bf B}\times\hat{{\bf k}})^{2}}{45m^4-8\alpha^2{\bf B}^2} \right]^{-1/2}
%\!\!\!\!\!\!
\simeq 1+ \frac{8 \alpha ^2}{45 m^4} ({\bf B} \times \hat{{\bf k}})^{2}
\; , \hspace{0.5cm}
\label{n1EH}
\\
n_{2}^{EH}
%\!\!\!&=&\!\!\!
%\left[1-\frac{28\alpha^2 \, ({\bf B}\times\hat{{\bf k}})^{2}}{45m^4+20\alpha^2{\bf B}^2} \right]^{-1/2}
%\!\!\!\!\!\!\!\!
\simeq 1+\frac{14 \alpha^2}{45 m^4} ({\bf B} \times \hat{{\bf k}})^{2}  \; . \hspace{0.5cm}
\label{n2EH}
\end{eqnarray}
\end{subequations}
It is worthy to highlight that these results are in agreement with Reference \cite{Adler71} after suitable changes in the unit system.
%
%These results are agreement with the result of Reference \cite{Adler71} after suitable changes in the unit system are done.
%

%
The second case of a ED non-linear is the generalized Born-Infeld (BI) Lagrangian \cite{Gaete2},
\begin{eqnarray}\label{BIg}
{\cal L}_{BI}({\cal F},{\cal G})= \beta ^2 \left[\, 1-\left(1-2\frac{{\cal F}}{\beta^2}-\frac{{\cal G}^2}{\beta ^4}\right)^p \, \right] \; ,
\end{eqnarray}
where $\beta$ is a scale parameter with dimension of squared energy (in natural units), and $p$ is a real parameter that satisfies $0<p<1$.
The usual Born-Infeld theory is obtained for $p=1/2$. %and taking $\beta^{-4} \rightarrow 0$. The Maxwell electrodynamics is recovered in the limit $\beta \rightarrow \infty$ and when $p=0.5$. Thus,
For $\beta \gg (|{\bf E}_{0}|,|{\bf B}_{0}|)$, the Lagrangian (\ref{BIg}) leads to % has the correction}
\begin{equation}
{\cal L}_{BI}\simeq 2 \, p \, {\cal F}+ \frac{1}{\beta^2} \left[ \phantom{\frac{1}{2}} \!\!\! 2p\left(1-p\right) \, {\cal F}^2 + p \, {\cal G}^2 \, \right]
%+O\left(\beta^{-3}\right)
\; . \; \;
\end{equation}
Here, we recall that Maxwell electrodynamics is recovered in the limit
$\beta \rightarrow \infty$, and $p=1/2$.
Before proceeding with the calculations of the dispersion relations, it is interesting to consider the electrostatic case,
with ${\bf B}_{0}=0$, the corresponding field equation in presence of charges is
\begin{eqnarray}\label{eqD}
\nabla\cdot{\bf D}_{0}=\rho \; ,
\end{eqnarray}
where $\rho$ denotes the charge density, and ${\bf D}_{0}$ is defined by
\begin{eqnarray}\label{D0E0}
{\bf D}_{0}=\frac{{\bf E}_{0}}{\left( 1- {\bf E}_{0}^{\, 2}/\beta^2 \right)^{1-p}} \; .
\end{eqnarray}
In the point-like particle case, $\rho({\bf r})=Q \, \delta^{3}({\bf r})$,  equation (\ref{eqD}) yields
${\bf D}_{0}=\hat{{\bf r}}\,Q/r^2$.
%with $Q=e/4\pi$.
So, the solutions of the electric field in (\ref{eqD}) are
difficult to obtain due to the polynomial equation with degree $(1-p)^{-1}$. The case $p=3/4$ is chosen and the solution for the
electric field is \cite{Gaete2}
\begin{eqnarray}\label{E0}
{\bf E}_{0}=\frac{\sqrt{3} \, \beta \, Q \, \hat{{\bf r}} }{\sqrt{Q^2+\sqrt{ Q^4+9\beta^4r^8 } }} \; .
\end{eqnarray}
The magnitude of the electrostatic field goes to zero whenever $r \rightarrow \infty$. In the limit $r \rightarrow 0$,
the electric field is finite at the origin, {\it i. e.}, $E_{0}(r=0)=\sqrt{3/2}\, \beta \, \mbox{sgn}(Q)$,
in which $\mbox{sgn}(Q)$ denotes the signal function. If the charge is positive, the electric field does not blow up at the charge position and the
maximum value it reaches is $\left.E_{0}\right|_{max}=\sqrt{3/2}\, \beta$. Otherwise, if the charge is negative, the electric field has
a minimum at $\left.E_{0}\right|_{min}=-\sqrt{3/2}\, \beta$. The electric potential associated with (\ref{E0}) is given by
\begin{eqnarray}\label{VBI}
V_{BI}(r)=\frac{3Q}{4(27)^{1/4}} \sqrt{\frac{\beta}{2Q}}
\left\{ \frac{64\sqrt{2\pi}}{15} \frac{\Gamma(9/4)\cos(\pi/8)}{\Gamma(3/4)}
\right.
\nonumber \\
\left.
+\frac{4}{3} \left[ \, \frac{\sin\phi_0 \, \cos\phi_0}{\cos^2\left(\phi_0/2\right)} \, \right]^{1/4}
\!\!-\frac{16}{3} \left[ \, \frac{\sin\phi_0}{\cos^2\left(\phi_0/2\right)} \, \right]^{1/4}
\times
\right.
\nonumber \\
\left.
\times \,
_2F_1\left( \frac{1}{8} , \frac{3}{4} , \frac{9}{8} , \tan^2\frac{\phi_0}{2} \right)  \right\} \; ,
\hspace{0.5cm}
\end{eqnarray}
in which $\phi_0=\tan^{-1}(3\beta^2r^4/Q^2)$. In addition, this potential reduces to the Maxwellian case for $\beta\rightarrow\infty$.
Using the result (\ref{VBI}), the electrostatic potential of an electron, with $Q=e=0.085$, evaluated at the origin $r\rightarrow0$
is finite given by
\begin{eqnarray}
\lim_{r \rightarrow 0} V_{BI}(r) = 0.62 \, \sqrt{\beta} \; .
\end{eqnarray}
%

%is recovered in (\ref{VBI}).
%The electrostatic energy stored in the like point-particle electric field is
%
%\begin{eqnarray}
%U_{BI}=\frac{1}{2} \int {\bf D}_{0} \cdot {\bf E}_{0} \, \, d^3{\bf r} \; ,
%\end{eqnarray}
%
%and using the results (\ref{D0E0}) and (\ref{E0}), we obtain
%
%\begin{eqnarray}
%U_{BI}=?? \; ,
%\end{eqnarray}
%

%
%
%We apply the expansion $F^{\mu\nu}=f^{\mu\nu}+F_{B}^{\,\,\,\,\mu\nu}$ up to second order at the $f^{\mu\nu}$ field strength tensor.
Now, returning to the dispersion relation analysis, the derivations from (\ref{coefficients}) 
in the Lagrangian (\ref{BIg}) yield the coefficients $c_{1}$, $d_{1}$ and $d_{2}$ for the generalized BI theory:
\begin{eqnarray}
\left. c_{1}^{BI} \right|_{{\bf E}=0,{\bf B}}&=&\frac{2p}{(1+{\bf B}^2/\beta^2)^{1-p}}
\; , \;
\nonumber \\
%\left.d_{1}^{BI} \right|_{{\bf E}=0,{\bf B}}&=&\frac{p(1-p)}{\beta^{2}(1+{\bf B}^2/\beta^2)^{1-p}} \; ,
 \left.d_{1}^{BI} \right|_{{\bf E}=0,{\bf B}}&=& \frac{4p(1-p)}{\beta^{2}(1+{\bf B}^2/\beta^2)^{2-p}} \; ,
 \nonumber \\
\left. d_{2}^{BI} \right|_{{\bf E}=0,{\bf B}}&=&\frac{2p}{\beta^{2}(1+{\bf B}^2/\beta^2)^{1-p}} \; .
\end{eqnarray}
The dispersion relations in this case are cast below:
\begin{subequations}
\begin{eqnarray}\label{reldispBI}
\omega_1^{BI}({\bf k}) \!&=&\! |{\bf k}| \, \, \sqrt{1 - 2(1-p) \, \frac{({\bf B}\times\hat{{\bf k}})^{2}}{ {\bf B}^2 + \beta^2}} \; ,
%\omega_1^{BI}({\bf k}) \!&=&\! |{\bf k}| \, \, \sqrt{1- (1-p) \, \frac{({\bf B}\times\hat{{\bf k}})^{2}}{2\beta^2}} \; ,
\label{omega1BI}
\\
\omega_2^{BI}({\bf k}) \!&=&\! |{\bf k}| \, \, \sqrt{1- \frac{({\bf B}\times\hat{{\bf k}})^{2}}{{\bf B}^2+\beta^2}} \; .
\label{omega2BI}
\end{eqnarray}
\end{subequations}
%
%that are real if the conditions $\beta^2+({\bf B}\cdot\hat{{\bf k}})^2>0$ and $2\beta^2>(1-p)({\bf B}\times\hat{{\bf k}})^{2}$ are respectively fulfilled.

\begin{figure}[t]
%\vspace{-5pt}
\centering
\includegraphics[width=0.47\textwidth]{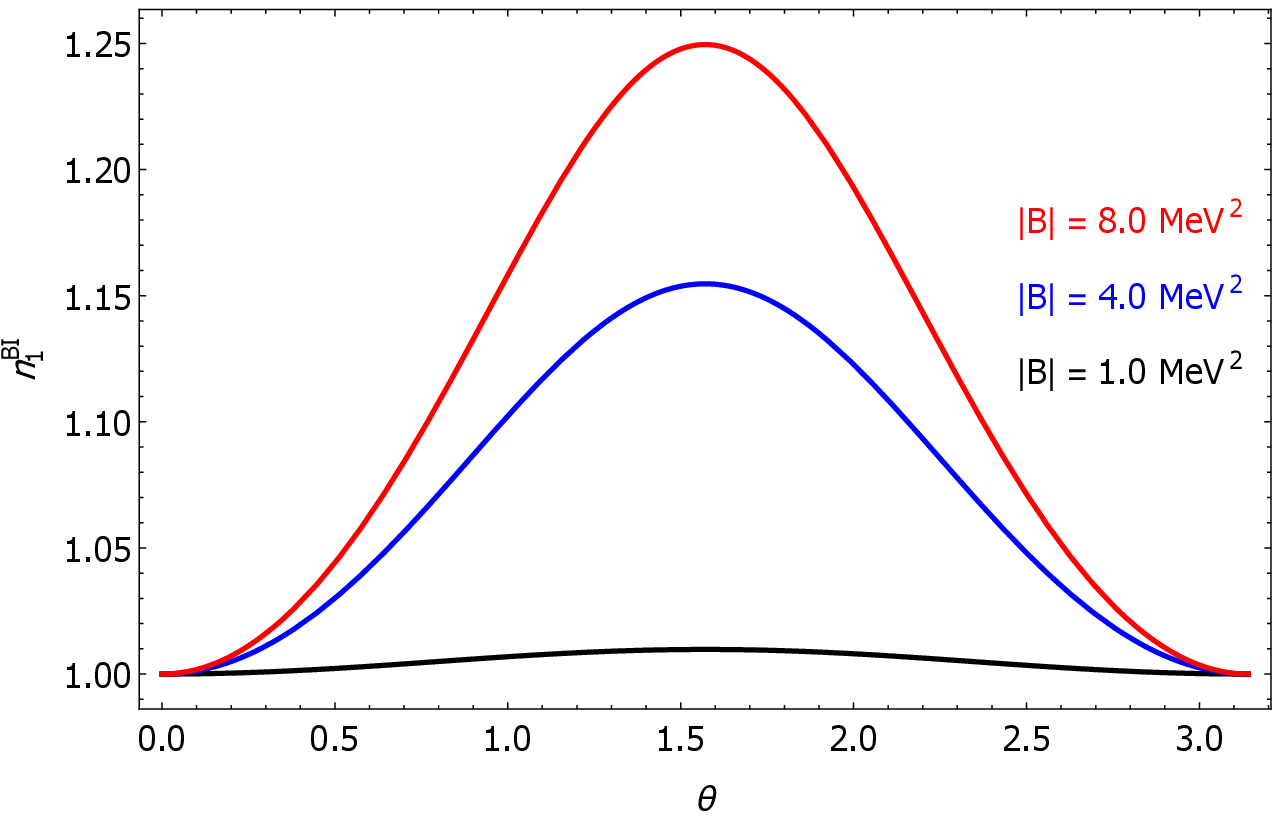}
\includegraphics[width=0.47\textwidth]{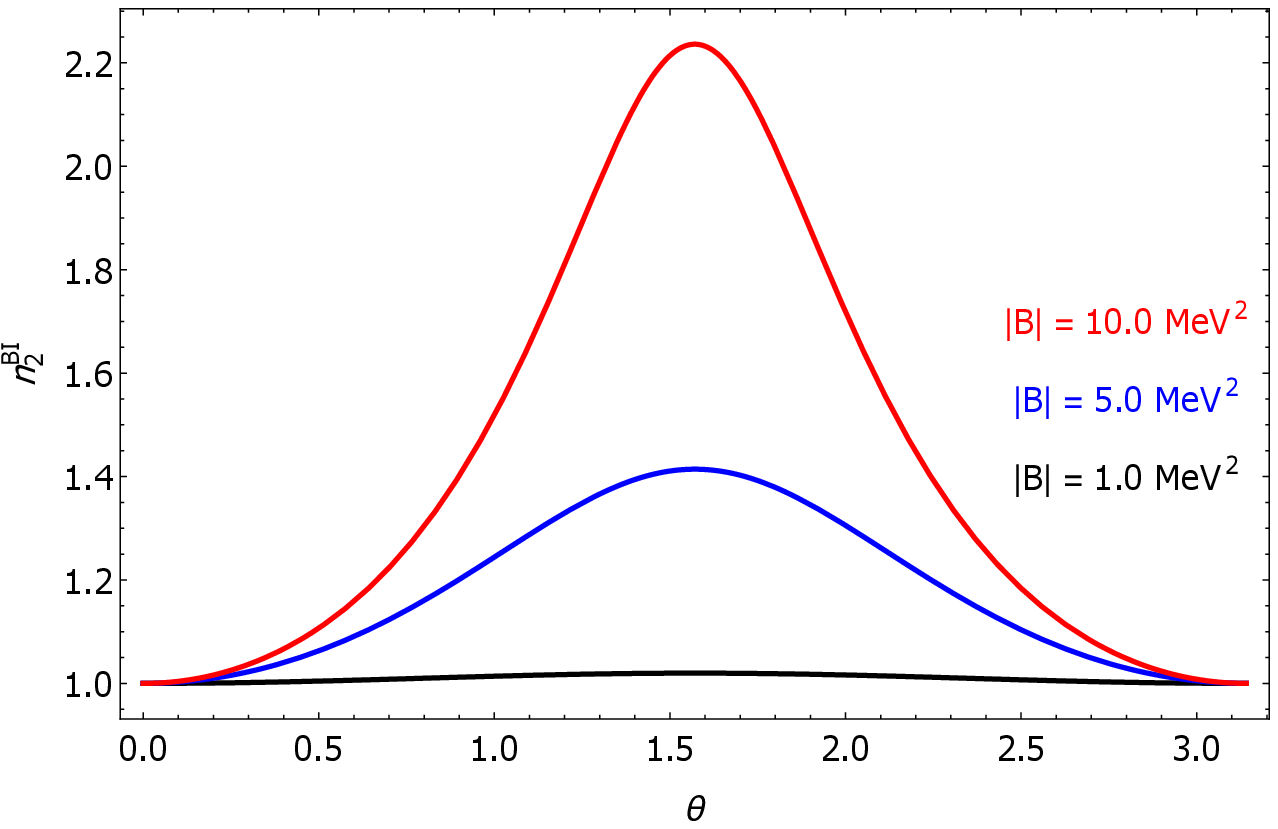}
\caption{The top panel : The refraction index of the
generalized Born-Infeld theory (\ref{omega1BI}) as function of the angle $\theta$
(between ${\bf B}$ and $\hat{{\bf k}}$), when $\beta=5.0\,\mbox{MeV}^2$ and $p=3/4$, for the values of
$|{\bf B}|=1.0\,\mbox{MeV}^2$, $|{\bf B}|=4.0\,\mbox{MeV}^2$ and $|{\bf B}|=8.0\,\mbox{MeV}^2$. The
bottom panel : The refraction index of the Born-Infeld theory (\ref{omega2BI}) for $|{\bf B}|=1.0\,\mbox{MeV}^2$, $|{\bf B}|=5.0\,\mbox{MeV}^2$
and $|{\bf B}|=10.0\,\mbox{MeV}^2$, when $\beta=5.0\,\mbox{MeV}^2$.}\label{ntheta}
\end{figure}

\begin{figure}[t]
%\vspace{-5pt}
\centering
\includegraphics[width=0.48\textwidth]{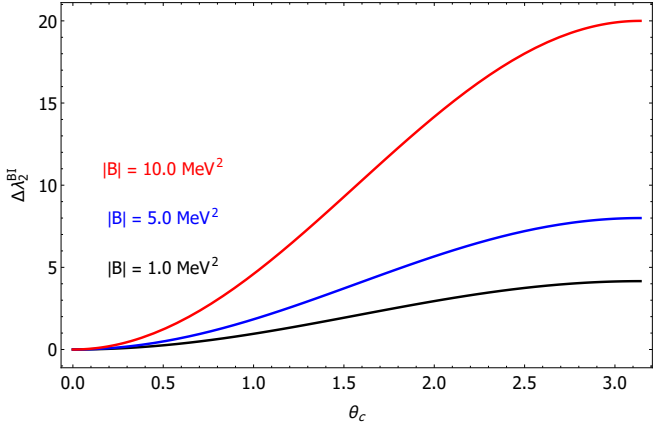}
\caption{The Compton effect in the usual Born-Infeld theory $(p=1/2)$.
The variation of the photon wavelength as function of the Compton angle, $\theta_c$, for the values of
$|{\bf B}|=1.0\,\mbox{MeV}^2$, $|{\bf B}|=5.0\,\mbox{MeV}^2$ and $|{\bf B}|=10.0\,\mbox{MeV}^2$.
We choose here $\beta=5.0 \, \mbox{MeV}^2$.
}\label{comptonBI}
\end{figure}

Notice that the second frequency is always real (since $\beta^2+({\bf B}\cdot\hat{{\bf k}})^2>0$) and independent of the $p$-parameter of the generalized BI theory. On the other hand, we need to be more carefully with the first frequency, which is real when
\begin{equation}
\beta^2 + (2p-1) {\bf B}^2 + 2(1-p) ({\bf B} \cdot \hat{{\bf k}})^{2} >0  \, \; .
\end{equation}
For the values $1/2 \leq p< 1$, this condition is fulfilled. However, for $0<p< 1/2$, will be necessary to impose some constraints between $\beta$ and ${\bf B}$. In addition, for the particular case of the BI theory $(p=1/2)$, we obtain $\omega_1^{BI} = \omega_2^{BI}$ and this frequency  leads to the well-known result in the literature of non-birefringence \cite{Birula2}. Whenever the magnetic field is strong, that is $|{\bf B}| \gg \beta$, the first frequency depends on the $p$ and the $\theta$ angle $\omega_{1}^{BI} \simeq |{\bf k}| \, \sqrt{p+(1-p)\cos(2\theta)}$ and the correspondent
refraction index is $n_{1}^{BI}\simeq [ \, p+(1-p)\cos(2\theta) \, ]^{-1/2}$, where $\theta$ is the angle between ${\bf B}$ and the direction of $\hat{{\bf k}}$. The second frequency for $|{\bf B}| \gg \beta$ is $\omega_{2}^{BI} \simeq |{\bf k}| |\cos\theta|$, and the refraction index is $n_{2}^{BI} \simeq |\sec\theta|$. Thereby, when the medium is under a strong magnetic background, the refraction index changes with the angle $\theta$ and it does not depend on the magnitude of the magnetic field.  Notice that the result of $\omega_{2}^{BI}$, for $|{\bf B}| \gg \beta$, coincides with the dispersion relation of \cite{SorokinPRD} (eq. 48) when the propagation is not superluminal.
%
%Also, it is noteworthy to point out that this result on the birrefringence is not in conflict with the fact that Born-Infeld Electrodynamics does not yield vacuum birrefringence. We are not denying this well-known fact. Our results cast above, in \eqref{omega1BI} and \eqref{omega2BI}, have not been derived in the the full-fledged Born-Infeld theory; they rather follow from the linearization of the latter, whenever the external magnetic field, ${\bf B}$, strongly dominates over the (propagating) electromagnetic excitations (the ${\bf e}-$ and ${\bf b}-$fields).}
%
%
The refraction index related to the generalized and usual BI theory as function of angle $\theta$ are plotted in the figure (\ref{ntheta}).
The top panel is the case of the generalized BI theory with $p=3/4$ and $\beta=5.0\,\mbox{MeV}^2$, for the background values $|{\bf B}|=8.0\,\mbox{MeV}^2$ 
(red line), $|{\bf B}|=4.0\,\mbox{MeV}^{2}$ (blue line) and $|{\bf B}|=1.0\,\mbox{MeV}^2$ (black line), respectively. The bottom panel is the usual BI theory 
$(p=1/2)$ with $\beta=5.0\,\mbox{MeV}^2$, for the background values $|{\bf B}|=10 \,\mbox{MeV}^2$ (red line), $|{\bf B}|=5.0\,\mbox{MeV}^{2}$ (blue line) 
and $|{\bf B}|=1.0\,\mbox{MeV}^2$ (black line), respectively, with $\beta=5.0\,\mbox{MeV}^2$. Notice that, in the limit $|{\bf B}| \rightarrow 0$, 
both the refraction index approach the value one.
The correction to the Compton effect (\ref{deltalambda}) for the usual BI theory $(p=1/2)$ is shown in figure (\ref{comptonBI}).
We plot the curves for the magnetic field values $|{\bf B}|=1.0\,\mbox{MeV}^2$ (black line),
$|{\bf B}|=5.0\,\mbox{MeV}^2$ (blue line) and $|{\bf B}|=10.0\,\mbox{MeV}^2$ (red line). The variation
of the photon wavelength in the Compton effect increases with the magnetic field magnitude.
%

\begin{figure}[t]
%\vspace{-5pt}
\centering
\includegraphics[width=0.49\textwidth]{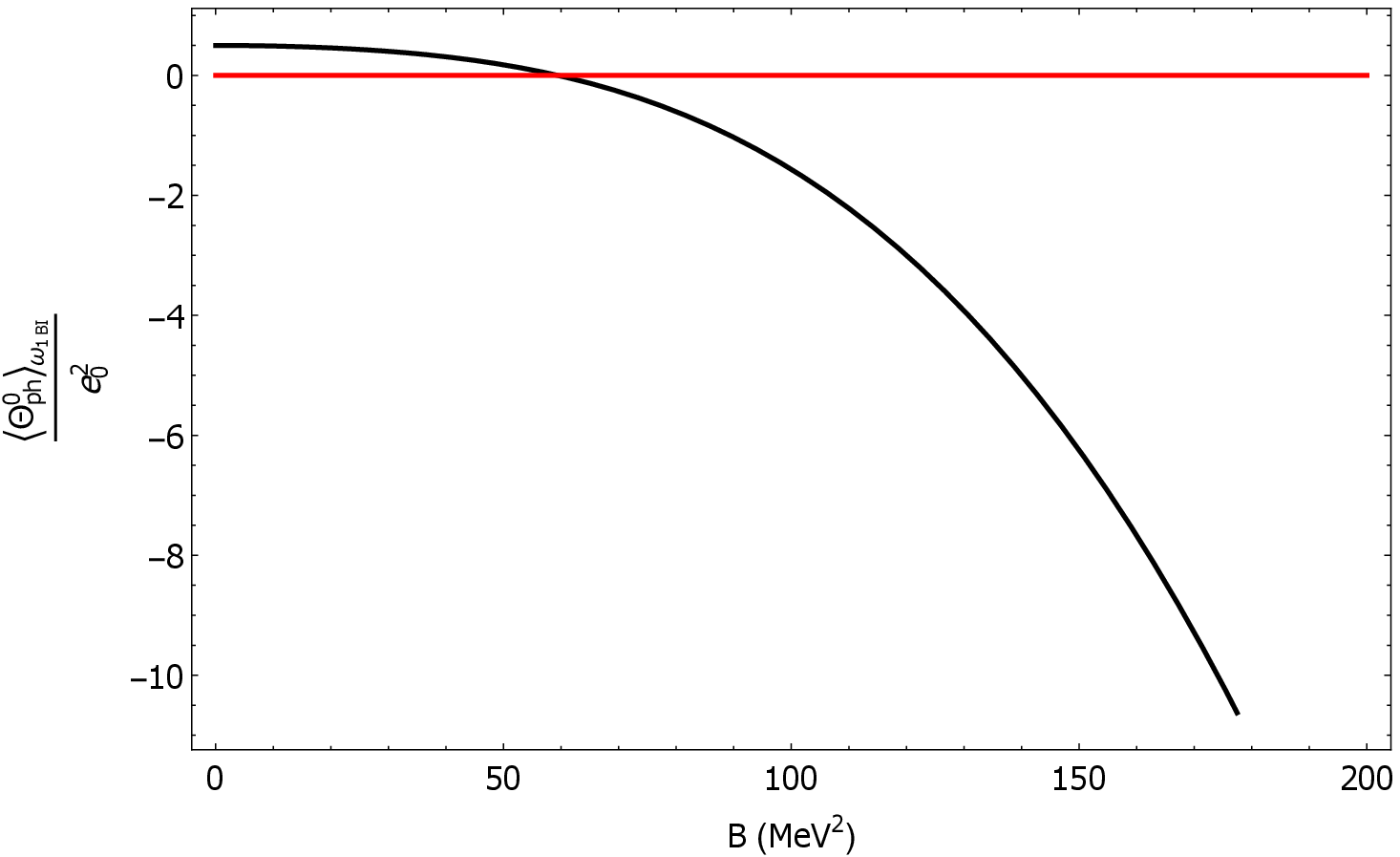}
\includegraphics[width=0.49\textwidth]{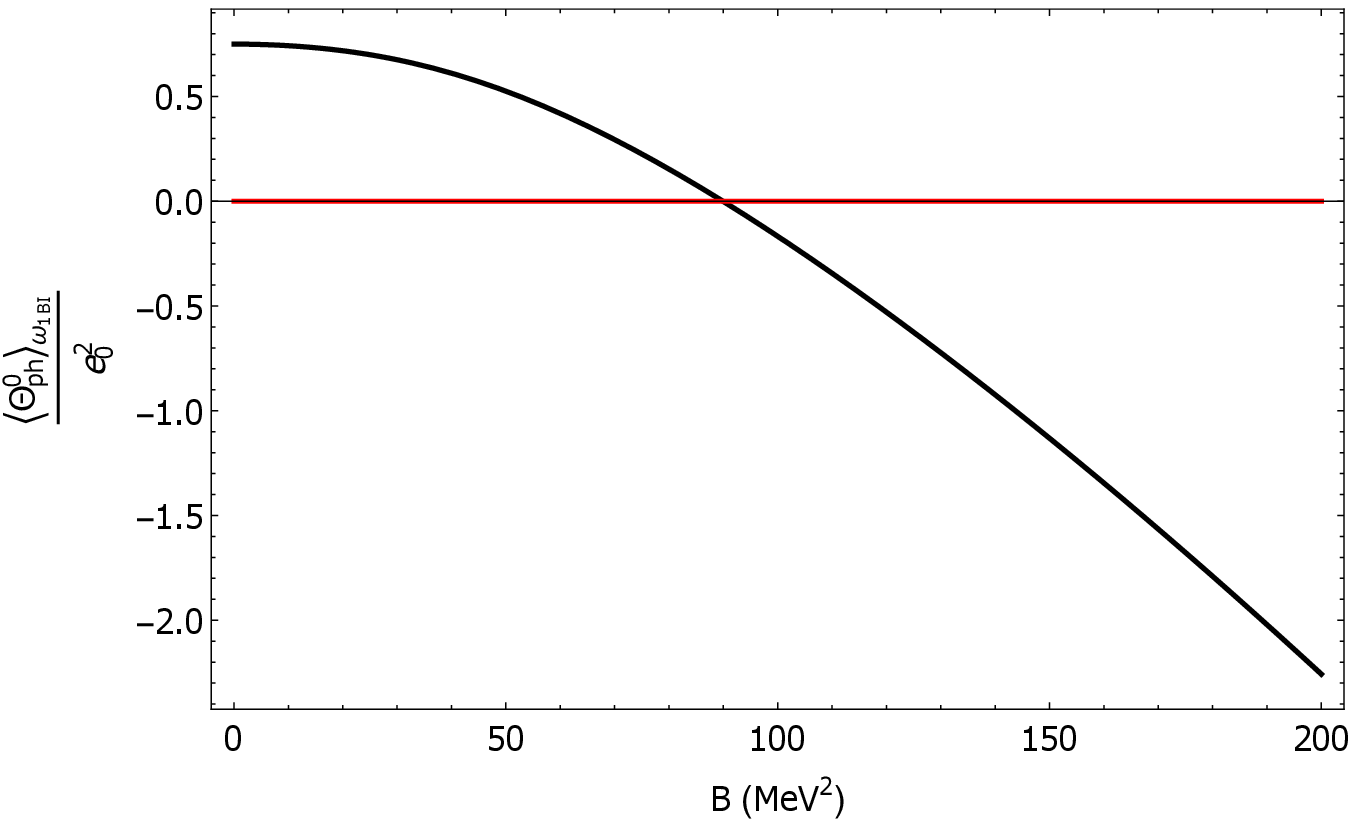}
\caption{The time average of energy density by unit of ${\bf e}_{0}^{2}$ for the usual BI ED $p=0.5$ (top panel),
and the generalized BI ED with $p=0.75$ (bottom panel) as function of the
background magnitude. We choose $\beta=50 \, \mbox{MeV}^2$ in this plot.}\label{thetaBI}
\end{figure}

%
%For the generalized BI ED, the energy density is positive if the magnetic background satisfies the condition $|{\bf B}| < \sqrt{2} \, \beta/\sqrt{1-p}$.
%
Using the plane wave for ${\bf e}$ and ${\bf b}$, the time average of the energy density in
generalized BI theory is given by
\begin{eqnarray}\label{Theta00BI}
\frac{\langle \Theta_{ph}^{00} \rangle^{BI} }{{\bf e}_{0}^{2}}=\frac{1}{4} \, c_{1}^{BI}\left[1+\frac{{\bf k}^2}{\omega^2}
-\frac{d_{1}^{BI}}{c_{1}^{BI}} \frac{{\bf k}^2}{\omega^2} \left( \hat{{\bf e}}_{0}\cdot(\hat{{\bf k}} \times {\bf B}) \right)^2
\right.
\nonumber \\
\left.
+\frac{d_2^{BI}}{c_{1}^{BI}} \left( \hat{{\bf e}}_{0} \cdot {\bf B} \right)^2
-\left(\frac{d_2^{BI}}{c_{1}^{BI}}\right)^2 \frac{{\bf k}^2}{\omega^2} \left( {\bf B} \cdot \hat{{\bf k}} \right)^{2}\left( {\bf B} \cdot \hat{{\bf e}}_{0} \right)^2
\right] \; . \; \; \;
\end{eqnarray}
%
%in which we have assumed the same conditions from the HI ED obtained in (\ref{Theta00HI}).
In this case, both frequencies
(\ref{omega1BI}) and (\ref{omega2BI}) depend on the external magnetic field, and just $\omega_{1}^{BI}$ depends on the $p$-parameter.
Therefore, the time average of energy density associated with the frequency $\omega_{1}^{BI}$ changes with the values of the
$p$-parameter. We plot the time average of energy density by unit of ${\bf e}_{0}^{2}$ associated with $\omega_{1}^{BI}$ for
$p=0.5$ (usual BI ED in the top panel) and $p=0.75$ (generalized BI ED in the bottom panel), when $\beta=50 \, \mbox{MeV}^2$,
in the figure (\ref{thetaBI}).
%{\color{red} (refazer os plot e a análise do campo crítico)}
%
In both plots, the time average of the the energy density becomes negative for $|{\bf B}|>|{\bf B}_{n}|$,
where $|{\bf B}_{n}|=59.46 \, \mbox{MeV}^{2}$ is a magnetic critical field for the case of the usual BI ED (top panel),
and $|{\bf B}_{n}|=89.94 \, \mbox{MeV}^{2}$ is a magnetic critical field for the case of the generalized BI ED (bottom panel).
Another interesting model is Logarithm ED \cite{Gaete}. The corresponding Lagrangian is given by
\begin{eqnarray}
{\cal L}_{ln}({\cal F},{\cal G})=-\beta^2 \, \ln \left[1-\frac{{\cal F}}{\beta^2}-\frac{{\cal G}^2}{2 \beta^4}\right] \; .
\end{eqnarray}
As in the previous case, Maxwell electrodynamics is recovered for $\beta \rightarrow \infty$.
In this model, we obtain the following coefficients $c_{1}$, $d_{1}$ and $d_{2}$ :
%In this case, the coefficients $c_{1}$, $d_{1}$ and $d_{2}$ are given by:
%
\begin{eqnarray}
\left. c_{1}^{ln} \right|_{{\bf E}=0,{\bf B}}&=&\frac{2\beta^2}{2\beta^2+{\bf B}^2}
\; , \;
\nonumber \\
\left.d_{1}^{ln} \right|_{{\bf E}=0,{\bf B}}&=&\frac{4\beta^2}{(2\beta^2+{\bf B}^2)^{2}} \; ,
\nonumber \\
\left. d_{2}^{ln} \right|_{{\bf E}=0,{\bf B}}&=&\frac{2}{2\beta^2+{\bf B}^2} \; .
\end{eqnarray}
Using these results, the combination of the coefficients $c_{1}^{ln}$, $d_{1}^{ln}$ and $d_{2}^{ln}$ in (\ref{omega2Bpm}) yields the same result (\ref{omega2BI}) for the second frequency in logarithm theory. The first solution (\ref{omega1Bpm}) in this case is
\begin{eqnarray}
\omega_{1}^{ln}({\bf k})=|{\bf k}| \, \sqrt{1-\frac{2|{\bf B} \times \hat{{\bf k}}|^2}{2\beta^2+{\bf B}^2}} \; .
\end{eqnarray}
For $|{\bf B}| \rightarrow \infty$, the correspondent refraction index is $n_{1}^{ln}=\sqrt{\sec(2\theta)}$.
The energy density is positive-definite if the magnetic background satisfies the condition $|{\bf B}| < \sqrt{2} \, \beta$.

\section{Conclusions and Final Remarks}
\label{sec6}
Our contribution sets out to pursue a study of general non-linear models of electrodynamics in presence of external electric and magnetic fields. 
The energy and linear momentum of the electromagnetic field fulfill a continuity equation if the background fields are uniform and constant. 
We have consider only the case of (uniform and constant) magnetic backgrounds for the analysis of the non-linear models we have picked out. 
The energy density is positive if the (external) field-dependent coefficients satisfy the conditions (\ref{condEp}). 
Otherwise, the energy density may assume negative values for sufficiently strong external fields, as our calculations point out. 
Plane wave solutions are considered to describe the non-linear photon. Two frequency solutions come out as it happens in the case of the usual photon; 
the other two frequencies exhibit a dependence on the uniform background magnetic field and the angle between the latter and the direction of the wave 
propagation. As a consequence, the refraction index also changes with the direction of the field ${\bf B}$ relative to the direction of propagation of the wave, $\hat{{\bf k}}$. 
We have also obtained the contribution of the uniform magnetic background to the kinematics of the Compton effect by employing the modified dispersion relations. 
We have applied these results to four examples of non-linear electrodynamics : Hoffmann-Infeld, generalized Born-Infeld, Logarithm and the Euler-Heisenberg effective Lagrangian. 
In all these cases, the refraction index depends on the angle, $\theta$, between ${\bf B}$ and $\hat{{\bf k}}$, and on the magnitude of ${\bf B}$ as well. 
The case of $|{\bf B}| \gg \beta$ is interesting because the results of the dispersion relation and refraction index are independent on the ${\bf B}$ magnitude. 
For the generalized Born-Infeld $n_{1}^{BI}\simeq [ \, p+(1-p)\cos(2\theta) \, ]^{-1/2}$, and for the usual Born-Infeld 
$n_{2}^{BI}\simeq|\sec\theta|=|\hat{{\bf B}}\cdot\hat{{\bf k}}|^{-1}$. Notice that $n_{1}^{BI}=n_{2}^{BI}$, when $p=1/2$. 
Furthemore, we have shown that the electrostatic potential is finite at the origin for $p=3/4$. 
The result for the Logarithm electrodynamics is $n_{1}^{ln}\simeq \sqrt{\sec(2\theta)}$, when $|{\bf B}| \gg \beta$.
%
%
%****************

%In the Euler-Heisenberg case, within the weak field approximation, the magnetic background has the upper bound $|{\bf B}| < 46.90 \, \mbox{MeV}^2$ for $n_{1}^{EH}$, and the upper bound $|{\bf B}| < 81.23 \, \mbox{MeV}^2$ for $n_{2}^{EH}$, which guarantees a real refraction index if $\hat{{\bf B}} \perp \hat{{\bf k}}$. Whenever $|{\bf B}|$ is very strong, the second refraction index, (\ref{n2EH}), in the EH theory is real if $\sin\theta < 0.84$.

%****************

%For end, we intend to study the contribution of the non-linear photon to the vacuum polarization in QED.
Further research has been initiated in a scenario involving models of scalar axions in connection with particular non-linear models. 
The purpose of this particular investigation is to try to understand how the present phenomenological astrophysical data known for the axions 
may dictate restrictions on the form of non-linear Lagrangian densities. On the other hand, we are also particularly interested in contemplating 
CP-breaking non-linear models in connection with the Planck 2018 Polarization Data to eventually use the latter to derive constraints on CP- violating 
non-linearities. We intend to report on the progress of these two lines of investigation in further papers.
Also, we would like to bring the reader's attention that the study we report in this contribution, mainly the investigation we carry out on dispersion relations, 
may be applied to extract bounds on the parameters of the non-linear models here inspected (and others we have not contemplated in this paper), 
if we consider current experiments, such as PVLAS and BMV, and the future high-power LASERs mentioned in the introduction, namely, SEL, ELI Project and XCELS. 
The work of Ref. \cite{Davila} raises the interesting possibility that large-scale LASERs may also be used in the 
LIGO, VIRGO and GEO interferometers to constrain the parameters of non-linear models by measuring the birefringence of the QED vacuum.

\section*{Acknowledgments}

\ni The authors express their gratitude to P. Gaete and A. Spallicci for stimulating discussions on diverse aspects of non-linear electrodynamic models.
F. Karbstein and D. Sorokin are also deeply acknowledged for pointing out important references and drawing our attention to relevant aspects of the Euler-Heisenberg and Born-Infeld Electrodynamics, respectively. M. J. Neves thanks CNPq (Conselho Nacional de Desenvolvimento Cient\' ifico e Tecnol\'ogico), Brazilian scientific support federal agency, for partial financial support, Grant number 313467/2018-8. L.P.R. Ospedal is grateful to the Ministry for Science, Technology and Innovations (MCTI) and CNPq for his Post-Doctoral Fellowship under the Institutional Qualification Program (PCI).

%
%\section{References}
%

\end{document}